\documentclass{emulateapj}

\def\gsim{\mathrel{\rlap{\lower 4pt \hbox{\hskip 1pt $\sim$}}\raise 1pt
\hbox {$>$}}}
\def\lsim{\mathrel{\rlap{\lower 4pt \hbox{\hskip 1pt $\sim$}}\raise 1pt
\hbox {$<$}}}

\begin{document}

\title{Properties of Newly Formed Dust Grains in The Luminous Type IIn Supernova 2010jl\altaffilmark{1}}
\shorttitle{Dust Formation in SN IIn 2010jl}
\shortauthors{Maeda et al.}

\author{
K.~Maeda\altaffilmark{2}, T.~Nozawa\altaffilmark{2}, D.K.~Sahu\altaffilmark{3}, Y.~Minowa\altaffilmark{4}, K.~Motohara\altaffilmark{5}, I.~Ueno\altaffilmark{6}, G.~Folatelli\altaffilmark{2}, T.-S.~Pyo\altaffilmark{4}, \\
Y.~Kitagawa\altaffilmark{5}, K.S.~Kawabata\altaffilmark{6}, G.C.~Anupama\altaffilmark{3}, T.~Kozasa\altaffilmark{7}, T.J.~Moriya\altaffilmark{2,8,9}, M.~Yamanaka\altaffilmark{6,10,11}, \\
K.~Nomoto\altaffilmark{2}, M.~Bersten\altaffilmark{2}, R.~Quimby\altaffilmark{2}, M.~Iye\altaffilmark{12}}

\altaffiltext{1}{Based on data collected at Subaru Telescope, which is operated by the National Astronomical Observatory of Japan, and at the Himalayan Chandra Telescope operated by the Indian Institute of Astrophysics.}
\altaffiltext{2}{Kavli Institute for the Physics and Mathematics of the 
Universe (WPI), Todai Institutes for Advanced Study, University of Tokyo, 
5-1-5 Kashiwanoha, Kashiwa, Chiba 277-8583, Japan; 
keiichi.maeda@ipmu.jp .}
\altaffiltext{3}{Indian Institute of Astrophysics, Koramangala, Bangalore 560034, India} 
\altaffiltext{4}{Subaru Telescope, National Astronomical Observatory of Japan, 650 North A'ohoku Place, Hilo, HI 96720, USA}
\altaffiltext{5}{Institute of Astronomy, University of Tokyo, 2-21-1 Osawa, Mitaka, Tokyo 181-0015, Japan}
\altaffiltext{6}{Hiroshima Astrophysical Science Center, Hiroshima University, 1-3-1 Kagamiyama, Higashi-Hiroshima, Hiroshima 739-8526, Japan}
\altaffiltext{7}{Department of Cosmosciences, Graduate School of Science, Hokkaido University, Sapporo 060-0810, Japan}
\altaffiltext{8}{Department of Astronomy, Graduate School of Science, University of Tokyo, 7-3-1 Hongo, Bunkyoku, Tokyo 113-0033, Japan}
\altaffiltext{9}{Reserch Center for the Early Universe, Graduate School of Science, University of Tokyo, 7-3-1 Hongo, Bunkyoku, Tokyo 113-0033, Japan}
\altaffiltext{10}{Department of Physical Science, Hiroshima University, 1-3-1 Kagamiyama, Higashi-Hiroshima, Hiroshima 739-8526, Japan}
\altaffiltext{11}{Department of Astronomy, Graduate School of Science, Kyoto University, Kyoto 606-8502, Japan}
\altaffiltext{12}{National Astronomical Observatory, Mitaka, Tokyo, Japan} 

\begin{abstract}
Supernovae (SNe) have been proposed to be the main production sites of dust grains in the Universe. Our knowledge on their importance to dust production is, however, limited by observationally poor constraints on the nature and amount of dust particles produced by individual SNe. In this paper, we present a spectrum covering optical through near-Infrared (NIR) light of the luminous Type IIn supernova (SN IIn) 2010jl around one and half years after the explosion. This unique data set reveals multiple signatures of newly formed dust particles. The NIR portion of the spectrum provides a rare example where thermal emission from newly formed hot dust grains is clearly detected. We determine the main population of the dust species to be carbon grains at a temperature of $\sim 1,350 - 1,450$K at this epoch. The mass of the dust grains is derived to be $\sim (7.5 - 8.5) \times 10^{-4} M_{\odot}$. Hydrogen emission lines show wavelength-dependent absorption, which provides a good estimate on the typical size of the newly formed dust grains ($\lsim 0.1 \micron$, and most likely $\lsim 0.01 \micron$). We attribute the dust grains to have been formed in a dense cooling shell as a result of a strong SN-circumstellar media (CSM) interaction. The dust grains occupy $\sim 10$\% of the emitting volume, suggesting an inhomogeneous, clumpy structure. The average CSM density is required to be $\gsim 3 \times 10^{7}$ cm$^{-3}$, corresponding to a mass loss rate of $\gsim 0.02 M_{\odot}$ yr$^{-1}$ (for a mass loss wind velocity of $\sim 100$ km s$^{-1}$). This strongly supports a scenario that SN 2010jl and probably other luminous SNe IIn are powered by strong interactions within very dense CSM, perhaps created by Luminous Blue Variable (LBV)-like eruptions within the last century before the explosion. 
\end{abstract}


\section{Introduction}

Core-collapse supernovae (CCSNe) have been suggested to be one of major production sites of cosmic dust grains. This has been highlighted by their capability of producing dust in the early Universe thanks to the short time scale of massive star evolution leading to core-collapse and explosion \citep[e.g., ][]{dwek2007}. SNe eject metal-rich materials into space, which cool rapidly and provide an ideal condition for dust condensation \citep{kozasa1989,todini2001,nozawa2003}. 

Another dust production mechanism has also been suggested  for a class of SNe exploding within dense circumstellar materials (CSM). Such SNe are classified as Type IIn due to their narrow hydrogen emission lines \citep{filippenko1997}, probably arising from unshocked CSM \citep{chevalier1994}, but also includes a rare class of SN Ib 2006jc-like objects \citep[Type Ibn: ][]{foley2007,pastorello2008, smith2008a, tominaga2008,anupama2009}. The strong interaction between the SN ejecta and CSM can provide an additional power source of the SN emission other than the radioactive decay chain $^{56}$Ni $\to$ $^{56}$Co $\to$ $^{56}$Fe \citep[e.g., ][]{chevalier1994,chugai2001,chugai2004,chevalier2011,chatzopoulos2012, moriya2012,moriya2013}. They show a range of luminosities \citep{taddia2013}, perhaps due to a diversity in the strength of the interaction (i.e., in the ejecta properties and/or the CSM density). In an extreme situation, either with a large kinetic energy of the explosion and/or extremely high density CSM, such a system may also be responsible for (at least a part of) Super-Luminous SNe \citep[SLSNe: see, ][and references therein]{quimby2011,galyam2012}. 

In these strongly interacting SNe, the dense shell would undergo strong radiative cooling, thus the temperature could decrease to the dust condensation temperature while the density is still high \citep{pozzo2004,smith2008a}. However, details of the process are yet to be clarified. One may expect that this `cooling-shell' dust formation takes place favorably in an earlier phase than in other classes of CCSNe as reported in a few cases \citep[e.g., ][]{smith2008a}, but constraining the dust formation time is generally difficult since it requires an extensive time-sequence from optical through IR to disentangle other possibilities like an echo from pre-existing dust grains in the CSM \citep[see, e.g., ][]{fox2011}. 

A main obstacle in understanding the importance of SNe in the cosmic dust inventory is generally insufficient data (or lack of detailed analysis) to specify the macroscopic and microscopic properties of newly formed dust grains in individual SNe. For most SNe suggested to have experienced dust formation, the link has been inferred by either a suppression of the red wing of spectral lines and/or a sudden drop in optical light curves \citep[see, ][for a review]{kozasa2009}. Stronger cases have been reported for some SNe showing an excess in IR wavelengths that is attributed to thermal emission from newly formed dust grains. For some SNe the spectral energy distribution (SED) was constructed from NIR and mid-IR photometry \citep[e.g., ][]{mattila2008,fox2009,andrews2010,fox2011}. The best cases have been provided by detection of a thermal component in their spectra, including SN II 1987A \citep[in Mid -- far-IR][]{moseley1989,wooden1993}, SN IIn 1998S \citep[in NIR:][]{gerardy2000,pozzo2004}, SN IIp 2004et \citep[Mid-IR:][]{kotak2009}, SN IIn 2005ip \citep[NIR -- Mid-IR:][]{fox2010}, and SN Ibn 2006jc \citep[NIR:][]{smith2008a,sakon2009}. 

The most convincing case has been reported for SN 1987A, where the dust is believed to have been formed deep in the SN ejecta through the concurrent appearance of the following events \citep[e.g., ][for a review]{kozasa1989}: (1) the blue shift in emission lines in optical (and NIR) \citep{lucy1989,meikle1993}, (2) decline in the optical luminosity and increase in the luminosities at longer wavelengths \citep{whitelock1989,suntzeff1990}, and (3) thermal continuum emission in mid-far IR \citep{moseley1989,wooden1993}. It is, however, very rare that all these three signatures are obtained for the same SN, given observational limitations. Typically, the spectral peak in the dust thermal emission is out of the observed wavelength range, and the quality of the spectra does not allow for a detailed investigation of line emissions at IR wavelengths. To the authors' knowledge, there has been no NIR spectroscopic observation of dust-forming SNe which covers a dust thermal emission peak and, at the same time, has the signal-to-noise (S/N) ratio sufficiently high to allow for a detailed study of line emissions. 

SN IIn 2010jl offers a new, promising site to study dust formation. This SN was discovered in an image taken on Nov. 3.52, 2010 (UT) \citep{newton2010} in nearby galaxy UGC5189A (redshift of 0.010697 and the luminosity distance of 48.9 Mpc, as we adopt throughout this paper, from the NASA/IPAC Extragalactic Database\footnote[13]{The NASA/IPAC Extragalactic Database(NED:  http://ned.ipac.caltech.edu/) is operated by the Jet Propulsion Laboratory, California Institute of Technology, under contract with the National Aeronautics and Space Administration}). It is classified as Type IIn \citep{benetti2010}, and is intrinsically bright, reaching an absolute magnitude of $\sim -20$ \citep{stoll2011,zhang2012}. Throughout this paper, we assume that the $V$-band maximum was reached at JD2455488 \citep{stoll2011} and we express the SN phase relative to this $V$-band maximum date. A pre-explosion image suggests that the progenitor of SN 2010jl was likely a massive star, with a main-sequence mass of $M_{\rm ms} \gsim 30 M_{\odot}$ \citep{smith2011}. Host galaxy properties imply that SN 2010jl exploded in a low metallicity environment, $Z \lsim 0.3 Z_{\odot}$ \citep{stoll2011}. SN IIn 2010jl was not only bright but also showed a slow evolution, similar to SN IIn 2006tf \citep{smith2008b}. While the initial decay rate was close to that of $^{56}$Co, later on around 100 days it flattened, and this flat evolution continued until at least 200 days \citep{zhang2012}. If the initial decay phase is interpreted as being powered by the decay of $^{56}$Co, it requires an unusually large amount of $^{56}$Ni synthesized at the explosion, as large as $3.4 M_{\odot}$ \citep{zhang2012}. Alternatively, the initial phase could be dominated by the SN-CSM interaction \citep{moriya2013}. In any case, there is no doubt that after $\sim 100$ days (as is the main focus of this paper) the radioactive energy input cannot be a major power source, and the SN-CSM interaction is the most likely power source. The total radiation energy output up to $\sim 200$ days in optical wavelengths, irrespective of the energy source, reached $\sim 4 \times 10^{50}$ ergs \citep{zhang2012}. The large luminosity and radiation output place SN 2010jl close to SLSNe \citep[][and references therein]{quimby2011,galyam2012}, thus understanding the properties of SN 2010jl could provide a hint in understanding yet-unresolved origins of these SLSNe. Strong SN-CSM interaction seems to provide a favorable site for the dust formation, perhaps through a dense cooling shell, as shown in some SNe IIn and the related SNe Ibn, including SN Ibn 2006jc, SNe IIn 1998S, 2005ip, and possibly SN IIn 2006tf and some other SNe IIn \citep{gerardy2000,pozzo2004,smith2008a,smith2008b,fox2010,fox2011,stritzinger2012}. 
This motivates our interest to study dust formation in luminous SN IIn 2010jl. 

Dust formation in SN 2010jl, at relatively early-phases until $\sim 200$ days after the discovery, has been under debate. IR excess peaking at $\sim 5 \micron$ was reported by \citet{andrews2011} on day $\sim 77$ (relative to V-band maximum). \citet{smith2012} reported systematic blueshift in hydrogen emission lines at similar epochs. While \citet{andrews2011} attributed the IR excess to an echo by pre-existing dust particles in the CSM, \citet{smith2012} speculated that these two features could indicate new dust formation. \citet{zhang2012} did not find a rapid decline in the optical light curve in the corresponding epochs, which is expected from the obscuration by newly formed dust particles, and suggested that the blueshifted emission lines might be due to a gas-opacity effect rather than the dust opacity \citep[see also ][]{smith2012}. 

In this paper, we present spectroscopic and photometric observations from optical through NIR of SN 2010jl at a late phase, about 550 days after the $V$-band maximum. We have found strong evidence for newly formed dust particles in this late phase, and our data set allows detailed study on the properties of the dust. In \S 2, we describe our observations and data reduction. Results are presented in \S 3, where we discuss properties of not only the dust grains but also the SN ejecta and CSM. The paper is closed in \S 4 with discussion and conclusions. An appendix is given for the optical spectrum data reduction, the SN environment, and for a constraint on a possible CSM echo in the SN optical spectrum.

\section{Observations and Data Reduction}

\subsection{NIR observations}

Our NIR observation was performed on April 24, 2012 (UT) (JD2456041.35) at the 8.2 m Subaru Telescope using the Infrared Camera and Spectrograph \citep[IRCS:][]{kobayashi2000} and the Subaru 188-element Adaptive Optics system \citep[AO188:][]{hayano2008,hayano2010}. The date corresponds to $+553$ days after the $V$-band maximum \citep{stoll2011}. For spectroscopy, we used $J$, $H$, and $K$ grism with a 52 mas pixel scale. The slit width was taken to be 0.225", resulting in a spectral resolution of $\sim 800$ in each band. Spectroscopy was performed in two different dithering positions A and B, with a total exposure time of 300, 600, and 1,200 seconds in $J$, $H$, and $K$, respectively. The air mass was $\sim 1$ for all observations. The SN was observed in the Laser Guide Star Adaptive Optics (LGS-AO) mode, where SN 2010jl itself was used as a Tip-Tilt star. An A0V star, HIP104353, was observed as a telluric standard star in the same night, in the same grism and slit setups with the Natural Guide Star (NGS) AO mode. The acquisition images were checked to confirm that the target center was correctly placed on a slit, thus the effect of a possible, artificial spectral warping is minimal \citep{goto2003}. 

We used IRAF for the spectroscopic data reduction.\footnote[14]{IRAF, the Image Reduction and Analysis Facility, is distributed by the National optical Astronomy Observatory, which is operated by the Association of Universities for Research in Astronomy (AURA), Inc., under cooperative agreement with the National Science Foundation (NSF).} We followed a standard procedure working on the difference images between two dithering positions, with flat fielding, bad pixel correction, and cosmic ray removal. The two-dimensional images were created for the negative and positive images with the standard star spectrum being a reference for the spectrum extraction. The wavelength was calibrated with comparison Ar lamp, which was then further checked with strong OH sky emissions. The one-dimensional spectra were extracted for the positive and negative images with further sky subtraction, where we optimized the aperture to provide the best signal-to-noise ratio without introducing an artificial change in the spectral shape. The positive and negative one-dimensional spectra were then combined to produce the final one-dimensional spectrum. The telluric absorption was removed using the standard star spectrum, where the A star atmospheric absorption lines had been removed before performing the telluric absorption correction. The flux calibration was also done using the standard star spectrum, assuming the black body temperature of 9,500K which we confirmed to be a good approximation in NIR for the Kurucz template A0V star spectrum. The flux was then further calibrated to match to the photometry (see below). 

The narrow field imaging observation was performed with the same instruments, with LGS-AO for SN 2010jl and NGS-AO for a standard star FS143 \citep{hawarden2001}. We used a 20 mas pixel scale, resulting in a Field of View (FoV) of $21"\times 21"$. Five-point dithering was performed with an exposure of 5 seconds in each position. Given the narrow field of IRCS, we could not perform relative photometry with field stars. Thus, we performed absolute photometry with the standard star FS143. The sky condition was good, but we note that a systematic error at the level of $\sim 0.1$ mag in the absolute magnitude scale might be involved (while the color in different bands would not contain such a large systematic error). The imaging data reduction was performed using IRAF, following standard procedures. Every image at a different dithering position was checked, and those with background fluctuations apparently different from the others were removed from the analysis (typically the first dithering position). Sky flat images were created with the standard star images, which were then used for flat fielding and sky subtraction. Photometry was then performed on the combined images from different dithering positions. We used an aperture extraction of the flux, where we varied the size of apertures to check the convergence. The FWHM without AO core was typically 0.6", and the typical size of the aperture we used was $\sim (3 - 5) \times$ FWHM. We confirmed that the zero-point we derived for each band  is consistent with the values reported in the instrument web page and past observations using the same instruments. We note that the standard star FS143 turns out to be a double star system spatially resolved in the AO-aided high resolution  $K'$-band image, thus we adopted a large aperture not to miss the total light from the system.  Our photometry results in $J = 14.09 \pm 0.1$ mag, $H = 12.89 \pm 0.04$ mag, and $K' = 11.98 \pm 0.02$ mag (where the errors account only for the statistical error).

\subsection{Optical observations}

In the optical region, SN 2010jl was observed with the 2m Himalayan Chandra Telescope (HCT) of Indian Astronomical Observatory, Hanle, India, on March 15, 2012 (JD 2456002.12) and May 20, 2012 (JD 2456068.18), corresponding to $+514$ and $+580$ days since the $V$-band maximum, respectively. The spectra  were obtained using  grisms Gr\#7 (wavelength range 3500 - 7800 \AA)  and Gr\#8 (wavelength range 5200 - 9200 \AA) available with the Himalaya Faint Object Spectrograph Camera (HFOSC). Arc lamp spectra of FeAr and FeNe for wavelength calibration and  spectra of spectrophotometric standard,  with broader slit to correct for instrumental response and  flux calibration, were also observed during the observing run.  Data reduction was carried out using 
various tasks available within IRAF. The images were bias-corrected and flat-fielded, then one dimensional spectra were extracted from the cleaned  image using the optimal extraction algorithm \citep{horne1986}. The extracted arc lamp spectra were used for wavelength calibration. The accuracy of the wavelength calibration was checked using the night sky emission lines and, whenever required, a small shift was applied to the observed spectrum.  The wavelength calibrated spectra were corrected for the instrumental response using the spectrum of spectrophotometric standards observed on the same night, and the supernova spectra were brought to a relative flux scale. The flux calibrated spectra in the two regions were combined to a weighted mean to obtain the final spectrum on a relative flux scale.  

\begin{figure}
\vspace{-1cm}
\hspace{-4.5cm}
        \begin{minipage}[]{0.95\textwidth}
                \epsscale{0.8}
                \plotone{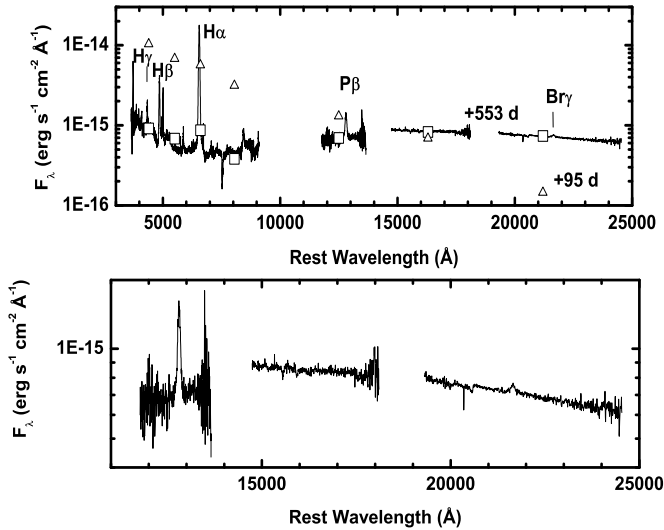}
        \end{minipage}
\vspace{-1cm}
\caption
{The combined spectrum of SN 2010jl at $+553$ days since the $V$-band maximum. The flux in each band converted from the photometry is shown by open squares for day + 553 (see main text). The early phase photometry is shown by open triangles, taken from \citet{zhang2012} (day $+87$) and \citet{andrews2011} (day $+95$). In the lower panel, we show only the NIR portion of the spectrum with a different scale in the vertical axis (for the sake of demonstration). 
\label{fig1}}
\end{figure}

Imaging observations of SN 2010jl were performed on March 27, 2012 (UT) and on May 21 (UT). These were obtained in the Bessell BVRI bands with the HFOSC mounted on HCT. With a plate scale of 0.296 arcsec/pixel, the central 2kx2k region of the SITe CCD chip used for imaging observations,  covers 10'x10' of the sky. The images were reduced in the standard manner using IRAF. As the supernova is embedded inside the galaxy we opted for the profile fitting photometry using daophot available within IRAF. 
A set of secondary standard stars was calibrated in the supernova field on several nights under photometric sky conditions. Finally, the supernova magnitudes were calibrated with respect to the secondary standards. The photometry results in $B = 16.96 \pm 0.04$ mag, $V = 16.64 \pm 0.03$ mag, $R = 15.71 \pm 0.05$ mag, $I = 15.79 \pm 0.02$ mag on March 27, and $B = 17.27 \pm 0.04$ mag, $V = 17.02 \pm 0.04$ mag, $R = 16.00 \pm 0.06$ mag, $I = 16.16 \pm 0.05$ mag on May 21.

Note that the above photometry includes a possible contamination from an unresolved background region, thus it is likely an overestimate of the SN magnitude. According to \citet{zhang2012}, the host galaxy contamination is fainter than 18 mag within the 3" diameter of the Sloan Digital Sky Survey (SDSS). We have roughly checked how much the contamination could be in our photometry as follows. Assuming that the {\em brightest} region within the host galaxy as a hypothesized unresolved background region, we estimate that the {\em upper limit} of the background contamination would be 0.6--0.8 mag in $B$, $V$, $I$, and 0.2--0.3 mag in $R$. More details on the light curve evolution and the estimate of the host galaxy contamination will be provided elsewhere (I. Ueno et al., in preparation). 

\begin{figure}
\hspace{-4.5cm}
        \begin{minipage}[]{0.95\textwidth}
                \epsscale{0.6}
                \plotone{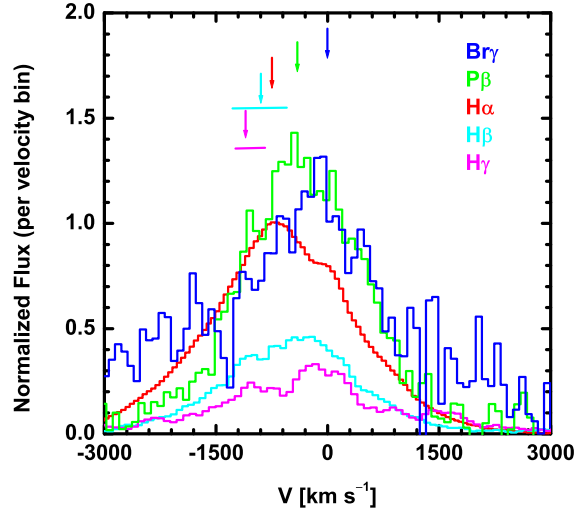}
        \end{minipage}
\vspace{-0cm}
\caption
{Comparison of the line profiles, shown for Br$_{\gamma}$ (blue), P$_{\beta}$ (green), H$_{\alpha}$ (red), H$_{\beta}$ (cyan), and H$_{\gamma}$ (magenta). The line flux is shown per velocity interval (i.e., $\Delta V \propto \Delta \lambda/\lambda_{0}$). The fluxes for the different lines are scaled according to the expected hydrogen emission line strength ratios in the Case B recombination \citep{osterbrock}, where the peak flux of H$_{\alpha}$ is taken to be unity for the normalization. Namely, if all the line strengths follow the Case B recombination, then all the lines should show the peak flux at about unity. To guide eyes, the rough positions of the central wavelength of the intermediate-width components are marked by arrows (with the same color coordinates with the corresponding lines). 
\label{fig2}}
\end{figure}

\section{Results}

In this paper, we argue that dust grains were newly formed in SN 2010jl, probably around 1 year after the $V$-band maximum. The case for the dust formation is regarded to be solid when the following three signatures are simultaneously detected \citep[see, e.g., ][for a review]{kozasa2009}: (1) Thermal emission from the dust grains, which was absent before the dust formation, (2) blueshift of emission line profiles, especially with the degree of the shift depending on the wavelength, and (3) decrease in the optical luminosity plus increase in the luminosity at longer wavelengths, corresponding to the thermal emission mentioned above. 

Typically, it is very rare that all these signatures are found for individual SNe (e.g., except for SN 1987A and possibly 2006jc: see references in \S 1). In case only one or two signatures are found, there is always a caveat that these features might be created by mechanisms other than the dust formation. This is the main difficulty in investigating the dust formation in SNe. Even with all these signatures detected, it is still possible to attribute each feature to be caused by different mechanisms, while it has been regarded that the new dust formation is the most natural (and straightforward) interpretation, since the dust formation scenario explains all these properties in a unified manner without fine-tuning. This is our philosophy in this paper -- since we have detected all these signatures in SN 2010jl, as we show in the following sections, we suggest that the case for dust formation is solid. We then analyze properties of the newly formed dust grains under the assumption that these features are mostly attributed to the formation of the dust. Further, we check this assumption (the dominance of the newly formed dust grains in causing these features) in view of other possible alternative mechanisms, and conclude that this assumption is well justified. 

\begin{figure}
\hspace{-4.5cm}
        \begin{minipage}[]{0.95\textwidth}
                \epsscale{0.6}
                \plotone{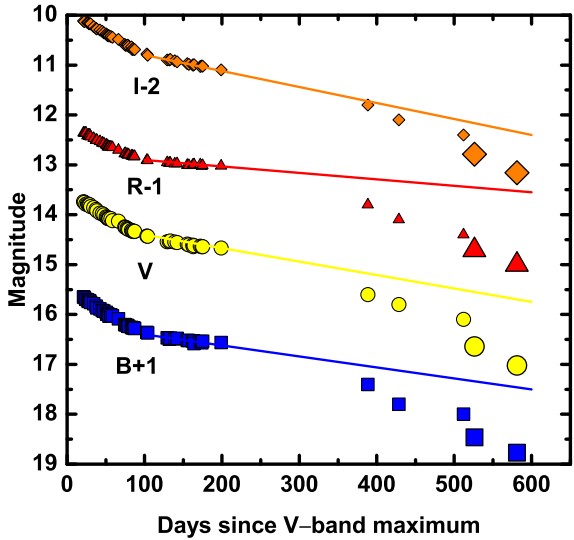}
        \end{minipage}
\vspace{-0.5cm}
\caption
{The optical light curves of SN 2010jl to late phases. The late-phase magnitudes obtained with the HCT telescope (large symbols) are combined with the earlier phase magnitudes reported in \citet{zhang2012}. The linear fits to the light curves in the phases between $+ 100$ and $+200$ days are shown by the solid lines \citep{zhang2012}. The late-phase magnitudes are below the extrapolation from the earlier phase (see \S 3.4 for details). Note that the unresolved background contamination has not been subtracted, which could reduce the intrinsic SN latest-phase magnitude further by $\lsim 0.6 - 0.8$ mag in $B$, $V$, $I$, and $\sim 0.2 - 0.3$ mag in $R$ (which correspond to the upper limit for the host contamination), while being negligible in the early phase. We suspect that the last point reported in \citet{zhang2012}, which does not agree with our photometry, would include a systematic error probably due to the background contamination; the difference may reflect the different size of the aperture in photometry. 
\label{fig3}}
\end{figure}

\subsection{Indications for the Newly Formed Dust Grains}

Figure 1 shows a combination of the IRCS spectrum and the HCT spectrum with the flux scaled to the date of the IRCS observation ($+553$ days). The spectrum is corrected for the redshift of the host galaxy ($z = 0.010697$) and for Galactic extinction from \citet{schlegel1998} ($E (B-V) = 0.027$ mag). No correction has been made for host-galaxy reddening, following \citet{smith2011}. For the optical spectrum, we adopted the one taken in March -- the spectral evolution during this phase was small, and for the purposes of this paper adopting the spectrum taken on May does not change our conclusions. In NIR, we have calibrated the flux scale with the photometry taken in the same night. In optical, we have first estimated the magnitudes at the same epoch by a linear interpolation between the photometry in March and May, then the flux scale is matched to this estimate. The evolution in the light curve is not large between these epochs ($\sim 0.3$ mag), and moreover the colors in the optical bands have stayed almost constant. As seen in Figure 1, the HCT spectrum matches the photometry at different bands very well. 

A striking feature in our spectrum is the clear detection of a thermal continuum component in NIR, which was absent in the early phase \citep[$\sim + 100$ days; ][]{andrews2011} as seen by comparing our late-time NIR spectrum and the SED at $\sim + 100$ days (Fig. 1). Unlike a few previous examples for which the continuum, increasing beyond the $K$-band, is detected in NIR spectra (see references in \S 1), the NIR emission peak in SN 2010jl is within the $H$ band, suggesting that the temperature of the emitting materials (irrespective of the nature of the material) is higher (i.e., $\sim 1,000 - 2,000$K) than the other examples. 

Besides the continuum, the NIR spectrum shows P$_{\beta}$ and Br$_{\gamma}$, but otherwise the spectrum is quite featureless. The optical spectrum exhibits Balmer series (H$_{\epsilon}$, H$_{\delta}$, H$_{\gamma}$, H$_{\alpha}$) and He I~$\lambda$5876, which all have intermediate-width components arising from the SN (see below). There are narrow emissions like [OIII]~$\lambda\lambda$4959,5007, but they are probably from an unresolved background region (see Appendix). All the resolved strong lines with the intermediate-width component are hydrogen lines (plus He I~$\lambda$5876), suggesting that the emitting region is either the shocked hydrogen-rich envelope or the hydrogen-rich CSM. 

Figure 2 shows line profiles of the hydrogen emission lines. The line flux is expressed per velocity interval (i.e., $\Delta V \propto \Delta \lambda/\lambda_{0}$, where $\lambda_{0}$ is the rest wavelength of a line). The fluxes of the different lines are scaled according to the expected hydrogen emission line strength ratios in the Case B recombination \citep{osterbrock} for the temperature of $\sim 5,000 - 10,000$K, where the peak flux of H$_{\alpha}$ is taken to be about unity for the normalization. Namely, if all the line strengths follow the expected emission line ratios for the Case B recombination, then all the lines should show the peak flux at about unity with this normalization. 

In optical wavelengths, the lines exhibit two components, a relatively narrow component at the rest wavelength (with the line width $\lsim 1,000$ km s$^{-1}$), and a broader (intermediate) component (FWHM $\sim 2,000$ km s$^{-1}$). The narrow component of H$_{\alpha}$ probably originates from the SN, as also seen in previous works \citep{smith2012,zhang2012}, while the narrow components in the other Balmer lines could be contaminated by the unresolved host and/or SN site (See Appendix).  Due to the limited spectral resolution in our observation, we are unable to distinguish these two possibilities from the kinematics \citep[$\sim 100$ km s$^{-1}$ or $28$ km s$^{-1}$ for the CSM case: ][]{smith2011}. In NIR, no strong narrow emission lines from the background are seen. Because of this possible contamination from the background region to the narrow component in the optical lines, we focus on the intermediate-width component hereafter. 

From Fig. 2 we see two clear features in the line profile: (a) the central wavelength of the intermediate-width component (see Fig. 2) is blueshifted (especially evident in the optical lines). Note that the degree of the blueshift is smaller for lines at longer wavelength, and with little or almost no blueshift for Br$_{\gamma}$. (b) A decrement in flux for lines at short wavelengths (especially evident in H$_{\beta}$ and H$_{\gamma}$) with respect to those expected for Case B recombination. Note that this decrement also persists if we normalize the line flux by the blue wing (\S 3.3), which means that at least a part of the decrement is independent of the normalization or intrinsic fluxes of different lines. 

It has been reported that the blueshift was already present in the early phase, and it was about $\sim 550$ km s$^{-1}$ between $100-200$ days \citep{smith2012,zhang2012}. We emphasize that the degree of the blueshift was larger in the late phase (after 400 days), reaching to $\gsim 700$ km s$^{-1}$ \citep[Fig.2: see also ][]{zhang2012}. Thus, it requires an additional mechanism to create (increase) the shift in the late phase, irrespective of a cause of the blueshift seen in the earlier phase \citep[that is still under debate: see ][]{smith2012,zhang2012}. This additional mechanism is typically attributed to the dust formation when other signatures of the dust formation are simultaneously detected. We also emphasize that our spectra show a wavelength-dependence in the degree of the blueshift, for which no mechanism has been suggested except for obscuration by dust particles. 

Figure 3 shows multi-band light curves of SN 2010jl including our photometry points. Between 100 and 200 days, SN 2010jl showed a flat light curve, which is interpreted as being powered by the SN-CSM interaction within smoothly distributed CSM materials \citep[e.g., ][]{zhang2012,moriya2013}. The light curve shows that the optical luminosity decline was accelerated around 1 year after the $V$-band maximum, and the luminosity at $\sim 550$ days is about 1 - 2 magnitudes (depending on the band and including uncertainty in the background contamination) below the extrapolation from the light curve at the earlier phases. 

As described above, we detect all three signatures of the formation of the dust grains in the late phase of SN 2010jl (see the first paragraph of \S 3). In the following sections, we will analyze the observational signatures and place constraints on properties of the newly formed dust grains. 

\begin{figure*}
\begin{center}
        \begin{minipage}[]{0.45\textwidth}
                \epsscale{1.0}
                \plotone{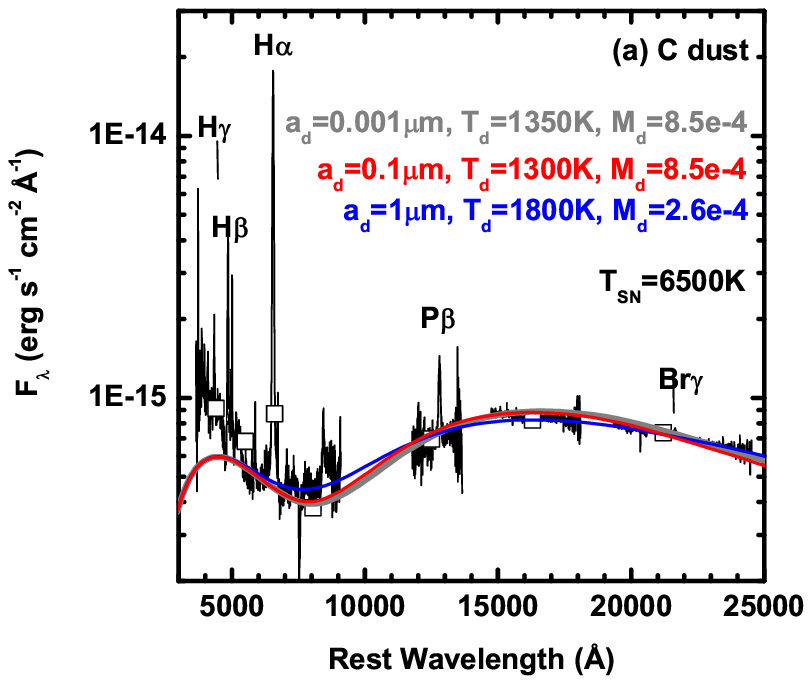}
        \end{minipage}
        \begin{minipage}[]{0.45\textwidth}
                \epsscale{1.0}
                \plotone{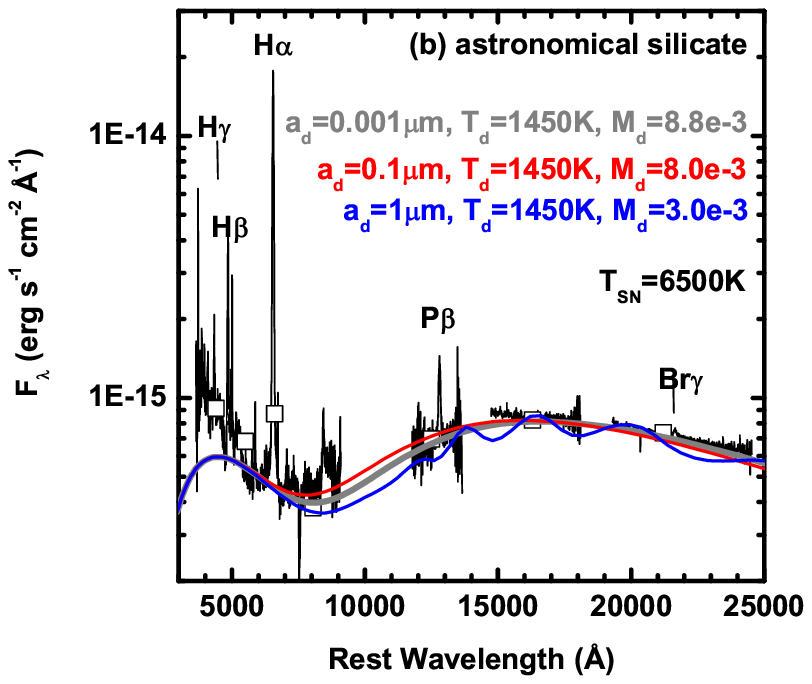}
        \end{minipage}
        \begin{minipage}[]{0.45\textwidth}
                \epsscale{1.0}
                \plotone{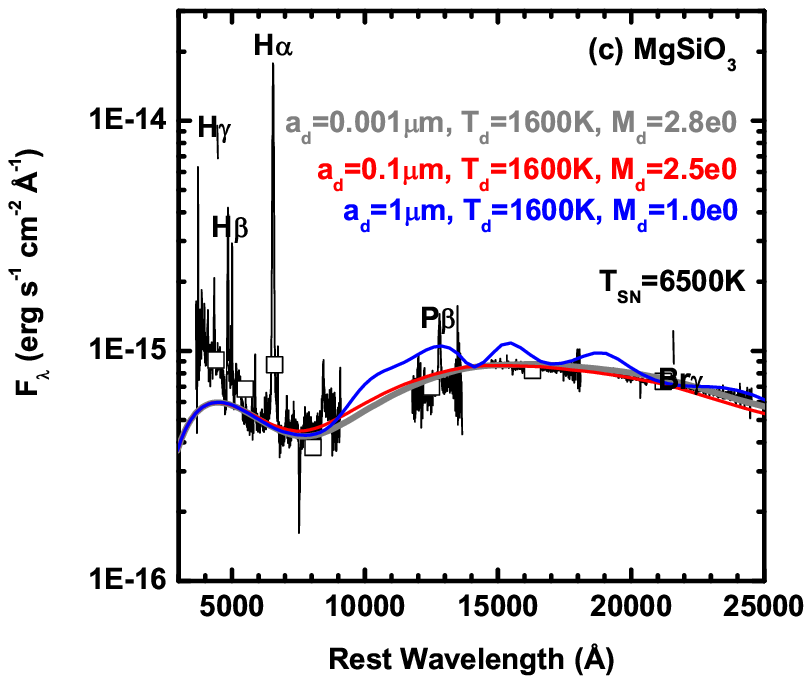}
        \end{minipage}
        \begin{minipage}[]{0.45\textwidth}
                \epsscale{1.0}
                \plotone{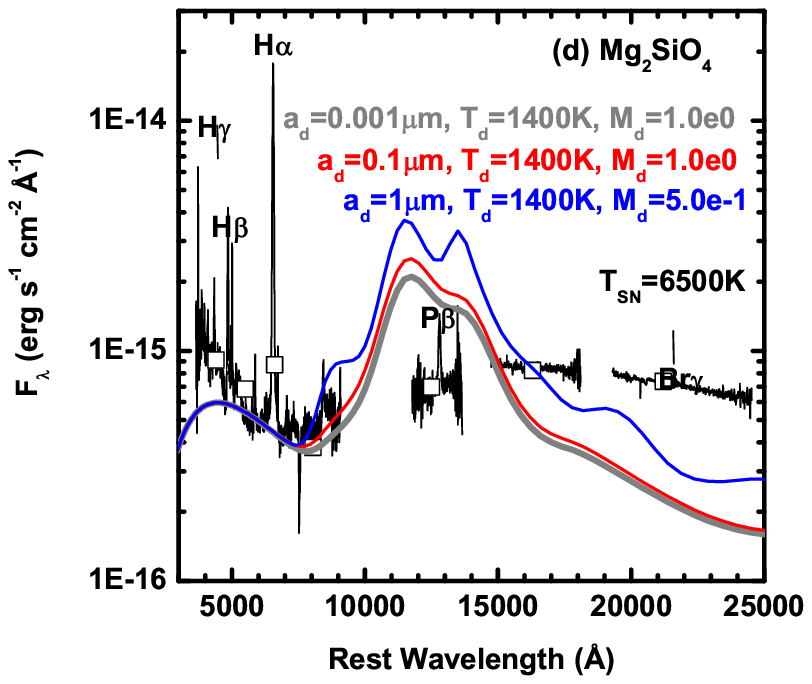}
        \end{minipage}
\end{center}
\vspace{-0.5cm}
\caption
{The observed spectrum as compared to the dust emission models for (a) carbon, (b) astronomical silicate, (c) MgSiO$_3$, and (d) Mg$_2$SiO$_4$ grains. The underlying SN emission is assumed to be a blackbody with the temperature of 6,500 K. Models with $a_{\rm d} = 0.01 \micron$ are identical to those with $a_{\rm d} = 0.001 \micron$, thus not shown here. 
\label{fig4}}
\end{figure*}

Our NIR spectrum is, to our knowledge, the first clear spectroscopic detection of thermal emission covering the peak from newly formed hot dust from an SN in NIR, and one of the strongest cases of dust formation in SNe to date -- the NIR thermal emission continuum from newly formed dust particles have been (spectroscopically) seen in a few SNe (see references in \S 1), but in these previous examples the emission peaks were out of the observed wavelength range. 

Since detecting the wavelength-dependent profile of hydrogen emission lines is not common due to observational limitations \citep[see, e.g., ][]{lucy1989,smith2008a,smith2012}, we here comment on this property. We note in particular that it is not only the blueshift but also the decrement that depends on the wavelength (Fig 2; see above). We suggest the following interpretation for this behavior: the intrinsic line profile of the intermediate-width component, for all the hydrogen lines, is at the rest wavelength but as one moves to shorter wavelengths, the intermediate-width component is blueshifted and suppressed due to presence of dust. This scenario explains the line profile change as a function of wavelength. The blueshift and suppression of the intermediate-width component are in line with the extinction by the newly formed dust particles as we see in the NIR continuum emission (see \S 3.2 and \S 3.3 for more details). The emission contribution in the red wing of the little-obscured NIR lines (i.e., P$_{\beta}$ and Br$_{\gamma}$) suggests that there is not a well-defined photosphere for the SN emission, and the SN emission is probably created as an optically thin emission in the interacting region (not as the photospheric emission arising below the interacting region). Otherwise, electron scattering would operate to suppress the emission from the receding side of the SN leading to the suppression of the red wing independent from the wavelength. This feature is difficult to reconcile with the obscuration created by the SN photosphere, as was discussed based on the early phase data \citep{smith2012,zhang2012}. The dust particles cannot be distributed deep in the SN ejecta, as in this case it would not produce the absorption at the rest wavelength \citep{smith2008a}. Thus, the dust particles should have a large overlap to the intermediate-width component emitting region in their distribution, supporting the formation of the dust particles in the cold-dense shell created at the interaction \citep{smith2008a}. We note that the narrow emission components in H$_{\alpha}$ and possibly in P$_{\beta}$ and Br$_{\gamma}$ do not show the wavelength shift, and this is consistent with the dust formation: these narrow emission features (while possibly contaminated by the background) are believed to arise from the unshocked CSM at a large distance, thus the newly formed dust grains within the ejecta or interacting region would not obscure much of the narrow emission components. 

We note that the suppression as compared to the expected line ratios from the Case B recombination should not be taken to be quantitative, since it is not clear if the Case B recombination is a good approximation in SN 2010jl. Indeed, the suppression of H$_{\gamma}$ and H$_{\beta}$ seems too large to be explained only by the extinction due to the newly formed dust particles, given the mild wavelength difference between these lines and H$_{\alpha}$. We note that the CSM density is probably very high in SN 2010jl (see \S 3.5), and it is very likely that even the first excited level of the neutral hydrogen is over populated. In this case, we expect a cascade of H$_{\gamma}$ and H$_{\beta}$ to H$_{\alpha}$ and other low-energy hydrogen series \citep{xu1992} like what happens for the Lyman series in the Case B case \citep{osterbrock}, thus the intrinsic ratio of H$_{\gamma}$ to H$_{\alpha}$ and that of H$_{\beta}$ to H$_{\alpha}$ may well be much smaller than the Case B recombination. This indeed supports the CSM interaction scenario for a power source of SN IIn 2010jl. We discuss more details on this CSM density in \S 3.5.

\subsection{Composition, Mass and temperature of the dust grains}

The NIR spectrum of SN 2010jl shows a clear thermal emission signature from newly formed dust particles. Figure 4 shows comparisons between the observed spectrum and the theoretically expected emission for different dust species and sizes. We adopt the blackbody temperature of 6,500 K for the underlying SN emission \citep{zhang2012} which fits well the SN emission in the optical wavelength at $\gsim 5,000$\AA. The blue continuum is not well fit by a single blackbody temperature: It may be contaminated by unresolved metal lines, e.g., Fe II \citep{foley2007}, or contaminated by a scattered optical light echo \citep{miller2010,andrews2011b}. As we will discuss in Appendix, the bluest potion of our spectrum is consistent with (but not exclusive to) the echo scenario, while the unresolved lines surely exist as well. We estimate that the echo would contribute at most $\sim 50$\% of the total flux in our HCT optical spectrum. 

The mass absorption coefficients of different dust species are calculated using optical constants by \citet{zubko1996} for amorphous carbon, \citet{draine2003} for astronomical silicate, \citet{dorschner1995} for MgSiO$_3$, and \citet{semenov2003} for Mg$_2$SiO$_4$. Graphite has the mass absorption coefficient similar to amorphous carbon in the wavelength range analyzed in this paper, thus we frequently refer the amorphous carbon and graphite grains as simply carbon dust. 

In this section, we assume that the dusty region is optically thin to the NIR photons (see \S 3.4 for a detailed discussion). The total mass of the dust grains ($M_{\rm d}$) and the temperature of the dust ($T_{\rm d}$) are connected by
\begin{equation}
F_{\nu} = \frac{M_{\rm d} B_{\nu} (T_{\rm d}) \kappa_{\nu}}{D^2} \ ,
\end{equation}
where $F_{\nu}$ is the flux of the dust thermal emission, $\kappa_{\nu}$ is the dust mass absorption coefficient (which depends on the dust species and the size distribution), and $D$ is the distance to the SN (48.9 Mpc). Since the thermal emission was spectroscopically detected in our NIR observation, we can obtain $M_{\rm d}$ and $T_{\rm d}$ without degeneracy for given dust species and the size. Our model fit is shown in Figure 4. 

In principle, with the temperature of the dust being a free parameter, the NIR thermal emission can be fit either by carbon, astronomical silicate, or MgSiO$_3$ grains. As long as the typical size of the dust ($a_{\rm d}$) is smaller than $\sim 0.1 \micron$, the size is not important in the NIR SED, and ($T_{\rm d}$, $M_{\rm d}$) $\sim$ ($1350$K, $8.5 \times 10^{-4} M_{\odot}$) for carbon and ($1450$K, $8.8 \times 10^{-3} M_{\odot}$) for astronomical silicate. If $a_{\rm d} \sim 0.1 \micron$, a slightly different set of ($T_{\rm d}$, $M_{\rm d}$) is required to fit the NIR spectrum, but the result is not very different from those for the dust with $a_{\rm d} < 0.1 \micron$. Dust particles as large as $a_{\rm d} \sim 1 \micron$ have a flat opacity dependence across the NIR wavelength range (as the dust size is similar to the electromagnetic wavelength), and a large value of $T_{\rm d}$ is required for carbon. The astronomical silicate grains with $a_{\rm d} \gsim 1 \micron$ do not fit the NIR spectrum irrespective of the temperature. Since the absorption coefficients of astronomical silicates are about an order of magnitude smaller than carbon, the dust mass ($M_{\rm d}$) required to fit the NIR spectrum is accordingly larger by about an order of magnitude for astronomical silicate than carbon. 

\begin{figure*}
\begin{center}
        \begin{minipage}[]{0.45\textwidth}
                \epsscale{1.0}
                \plotone{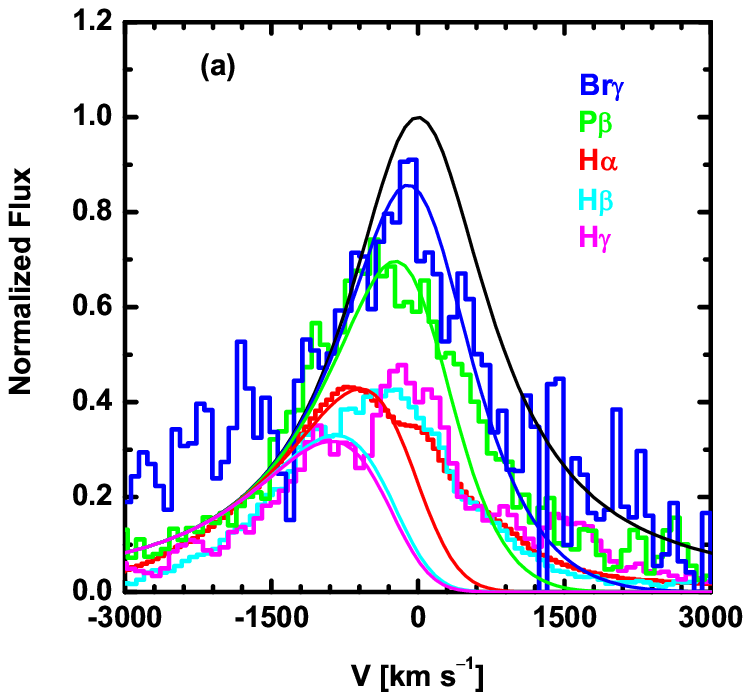}
        \end{minipage}
        \begin{minipage}[]{0.45\textwidth}
                \epsscale{1.0}
                \plotone{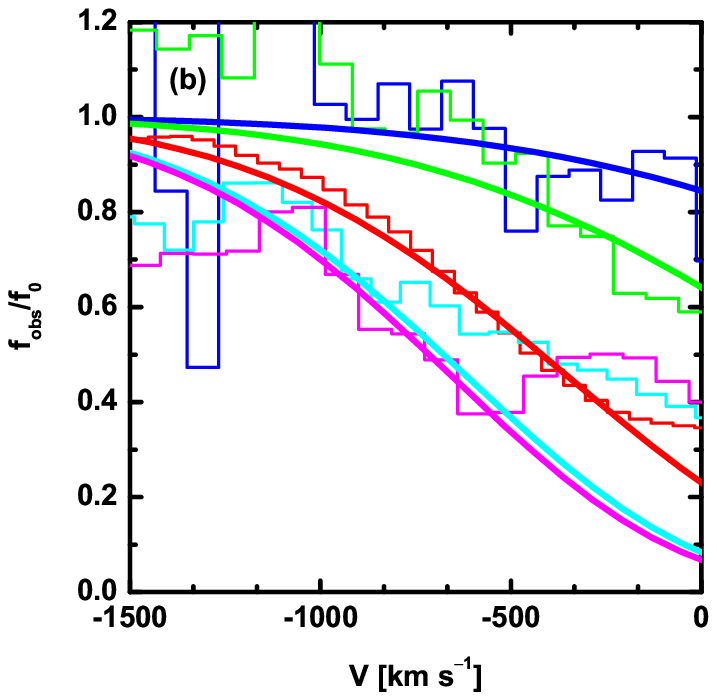}
        \end{minipage}
\end{center}
\vspace{-0.5cm}
\caption
{Interpretation of the hydrogen line profiles by obscuration through the newly formed dust grains. The normalization for the line flux is taken at the blue wing of the lines, i.e., assuming that the line profiles approach non-obscuration in the blue wing since this is the emission from the approaching side of the SN. (a) The black line is the assumed intrinsic line profile, a Lorentzian with FWHM of $1,800$ km s$^{-1}$. Different lines show the obscured line profiles with different amount of absorption characterized by $\tau_0$, taken to be $\tau_0 = $ 0.17 (Br$_{\gamma}$), 0.44 (P$_{\beta}$), 1.5 (H$_{\alpha}$), 2.5 (H$_{\beta}$), and 2.7 (H$_{\gamma}$). (b) Same but the flux scale normalized by the assumed intrinsic line profile, shown for the blue wing. This figure shows the escaping fraction of light as a function of the wavelength (i.e., the emitting position, which is closer to the observer along the line of sight for the light at shorter wavelength). The thick (smooth) curves are the same models as shown in (a). Note that our "obscured" model is intended to provide a reasonable fit only to the blue wing, and we are not aiming at fitting the red wing simultaneously (see \S 3.3). Also, the `narrow' component seen in the optical lines (H$_{\alpha}$, H$_{\beta}$, and H$_{\gamma}$) is fully omitted from our analysis, so the discrepancy around the rest wavelength is totally an artifact. 
\label{fig5}}
\end{figure*}

The fit by MgSiO$_3$ grains provides qualitatively similar results to those by astronomical silicate, while quantitatively there is a big difference. The wavelength dependence of the absorption coefficients of MgSiO$_3$ is even flatter than astronomical silicate in the NIR wavelength range, and thus $T_{\rm d}$ is larger for MgSiO$_3$. Moreover, the dust mass has to be as large as a few $M_{\odot}$ if the newly formed dust is MgSiO$_3$, due to the small absorption coefficients of MgSiO$_3$ in NIR. A similar situation is the case for Mg$_2$SiO$_4$, where $M_{\rm d} \sim M_{\odot}$ is required to explain the flux level of the NIR emission. Indeed, Mg$_2$SiO$_4$ predicts a characteristic bump in the $H$ band, which does not fit to the observed smooth thermal emission. 

In sum, either carbon or astronomical grains remain possible to reconcile the NIR emission property, if the dust temperature is simply taken as a free parameter. MgSiO$_3$ is highly unlikely, given the extremely large dust mass required to fit to the NIR emission. We exclude Mg$_2$SiO$_4$ solely on the basis of the unacceptable fit to the NIR emission spectrum. Additional information is given in other wavelength ranges, and in \S 3.5 we will show that the opacity behavior in mid-IR can also be used to reject MgSiO$_3$ and Mg$_2$SiO$_4$ even without direct  observations in such a wavelength range. 

The required temperature of the dust is too high for the hypothesized astronomical silicate dust and larger than the condensation temperature, which is typically $\sim 1,000$K \citep{nozawa2003}. This indicates that the newly formed dust is mainly carbon, either graphite or amorphous carbon. The condensation temperature for carbon grains is $\sim 1,900$K, fully consistent with the lower temperature we have derived. As analyzed above, the derived properties of the carbon dust are not sensitive to the unknown size of typical dust particles as long as $a_{\rm d} \lsim 0.1 \micron$, and these are $\sim 1,300 - 1,350$K and $\sim 8.5 \times 10^{-4} M_{\odot}$, while 
these are $\sim 1,800$K and $\sim 2.6 \times 10^{-4} M_{\odot}$ if $a_{\rm d} \sim 1 \micron$. 

\subsection{Size of the dust}

Constraining the size of the dust particles is generally a tough task for astronomical objects, one that requires deriving the extinction curve across a broad wavelength range. This is especially a severe problem in studying newly formed dust particles in SNe, as the intrinsic SED of SN emission is not known a priori. For SN 2010jl, we propose an approach to tackle this problem, by investigating the line profiles of hydrogen emission lines from optical through NIR, and by attributing the difference in the line profiles to different optical depths at different wavelength. 

As shown in Figure 2, the intermediate-width component of the hydrogen lines shows a blueshift especially evident in optical. This suggests that the line-emitting region and the dusty region have a large overlap, and the receding part of the emitting region is (partly) hidden by the newly formed dust particles (\S 3.1). The larger degree of the blueshift at shorter wavelengths also supports dust obscuration \citep{smith2008a}.

Now, under a reasonable assumption that all the intermediate-width component hydrogen lines are emitted from the same region, we can obtain a rough constraint on how the dust opacity depends on wavelength (i.e., the extinction curve). Under this assumption, the intrinsic line profile is the same for all the lines, and the photon path length within the dusty, obscuring region is identical for all the lines. Namely, we can connect the observed profile ($f (v)$) and the intrinsic line profile ($f_{0} (v)$) as follow, where $v/c \equiv \lambda/\lambda_0 - 1$ ($\lambda_0$ is the rest wavelength of the lines, and $c$ is the speed of light): 
\begin{equation}
f (v) = f_0 (v) \exp(-\tau_{\lambda} (v)) \ ,
\end{equation} 
where $\tau_\lambda (v) = \int \rho \kappa_{\lambda} ds \propto \kappa_{\lambda}$ ($\rho$ is the density of the dust particles and $ds$ is the path length, which are both independent from the wavelength). The function $\tau_\lambda (v)$ describes the amount of absorption for a photon emitted from a given position. While the optical depth ($\tau_{\lambda} (v)$) depends both on the opacity ($\kappa_{\lambda}$) and the column density, the latter is independent from the wavelength. Accordingly, the {\em difference} in the observed profiles of different lines comes only from different values of $\kappa_{\lambda}$ under the assumptions described above. Although we do not know the intrinsic strengths of the hydrogen lines from different transitions (which do not follow the Case B recombination; \S 3.1), we expect that the blue wing of the lines should approach non-obscuration, thus we scale the observed line profiles of different lines (i.e., $f (v)$) at the blue wing. 

\begin{figure}
\hspace{-4.5cm}
        \begin{minipage}[]{0.95\textwidth}
                \epsscale{0.6}
                \plotone{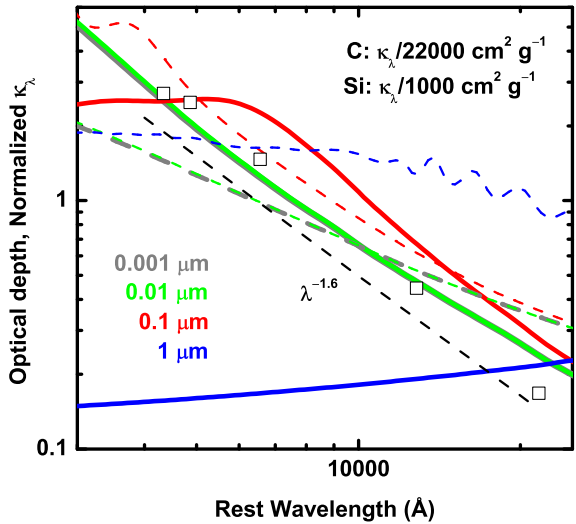}
        \end{minipage}
\vspace{-0.5cm}
\caption
{Wavelength dependence of the optical depth estimated from the emission lines, as compared with the theoretical expectations. The open squares are the observationally derived optical depths for different hydrogen emission lines. The dashed black line represents a power law behavior which fits to the optical depths at different wavelengths we have derived. The other lines are the theoretical expectations for carbon dust (thick-solid lines) and for astronomical silicate (thin dashed lines). The different size of the grains used are  shown with different colors ($a_{\rm d} = 0.001 \micron$ by gray, $0.01 \micron$ by green, $0.1 \micron$ by red, and $1 \micron$ by blue). For the theoretical curves, the opacities are normalized by 22,000 cm$^2$ g$^{-1}$ and $1,000$ cm$^{2}$ g$^{-1}$ for carbon and astronomical silicate, respectively. 
\label{fig6}}
\end{figure}

The functional form of the absorbing fraction as a function of the wavelength across the same line (i.e., dependence of $\tau_\lambda (v)$ on $v$) depends on the structure and kinematics of the emitting and absorbing regions, and is generally not expressed in an analytic form. Still, we can test an approximated simple form for it, by the requirement that the profiles of different lines can be at least qualitatively explained by varying the single parameter $\tau_{\lambda}$. We have tested simple polynomial forms, and in Figure 5 we show an example where we adopt $\tau_{\lambda} (v) = \tau_0 ((v/c + 0.01)/0.01)^\beta$ (i.e., transparent for $v/c < -0.01$) and $\beta=5$. This form provides a reasonably good description of the blue portions of different lines, with the intrinsic line profile described by a Lorentzian \citep[e.g., ][]{stritzinger2012}. The normalization of the optical depth, $\tau_0 (\lambda)$, is a parameter which is different for different lines and estimated by fitting roughly a position of the peak flux in the absorbed line profiles. The values for $\tau_0$ we adopt in Figure 5 is the following: $\tau_0 = $ 0.17 (Br$_{\gamma}$), 0.44 (P$_{\beta}$), 1.5 (H$_{\alpha}$), 2.5 (H$_{\beta}$), 2.7 (H$_{\gamma}$). The functional form we adopted above creates too much absorption in the red, but we are mainly concerned with the behavior in the blue to derive the relative optical depth for the different lines.  We are not aiming at fitting the detailed line profiles, since even without the absorption the interpretation of the intrinsic line profile in SNe IIn has not been fully clarified yet \citep[see, e.g., ][]{chugai2001,chugai2004}. We emphasize that our procedure here is focusing on deriving the optical depth difference at different wavelengths -- our treatment is quite general, and the results in the optical depth differences would not be sensitively dependent on the assumed functional form. This method allows the dependence of the opacity on the wavelength to be derived rather independently from the detailed kinematical properties of the emitting region.

In the right panel of Figure 5, we normalize the line profiles by the assumed intrinsic line profile, i.e., a Lorentzian with the FWHM of 1,800 km s$^{-1}$. This provides the escaping fraction of the light as a function of the wavelength (i.e., the path length within each line). The same analytic curves are shown in this plot, which shows that the decreasing escaping fraction from the blue to the red is well described by our formalism. 

With the estimate of the wavelength-dependent optical depth (which is only dependent on the opacity), we can extract a rough behavior of the extinction of the newly formed dust as a function of wavelength. In Figure 6, we compare the phenomenologically derived optical depth (for different hydrogen lines as shown by open squares) to the theoretical expectations for different dust species and grain sizes (for carbon and astronomical silicate). The dust model is the same as that used in Figure 4 for the thermal emission. The first thing to note is that the observationally derived optical depth ($\kappa_{\lambda} \propto \lambda^{-1.6}$) follows roughly the expected behavior for carbon dust ($\sim \lambda^{-1.4}$) when the dust size is smaller than the wavelength of interest. Given the uncertainties involved in our derivation, the behavior we have derived is surprisingly close to the theoretical expectation. It favors again carbon over silicate, independent from the argument of thermal emission, as the silicate dust predicts a flatter opacity dependence (note that MgSiO$_3$ has even flatter distribution than astronomical silicate).  Also, we do not see clear saturation of the opacity at shorter wavelengths, constraining the dust size as $a_{\rm d} \lsim 0.1 \micron$ (more likely $a_{\rm d} \lsim 0.01 \micron$ from the derived opacity difference within the optical wavelength). 

\begin{figure}
\hspace{-2.5cm}
        \begin{minipage}[]{0.7\textwidth}
                \epsscale{0.92}
                \plotone{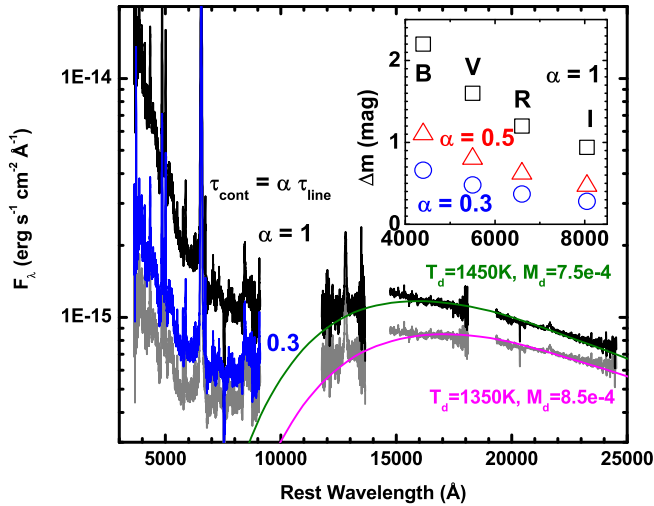}
        \end{minipage}
\vspace{-0.5cm}
\caption
{Possible effects of the dust absorption. The observed (gray) and the absorption-corrected (black) spectra are shown (assuming $\alpha = 1$ in the optical wavelength; $\tau_{\rm cont} = \alpha \tau_{\rm line}$, where $\tau_{\rm line}$ was derived independently). In optical, an intermediate case with $\alpha = 0.3$ is also shown (blue). NIR thermal emission models are shown for the observed (magenta) and absorption-corrected (green) spectra. The inset shows the expected magnitude decrease due to the internal dust absorption, for $\alpha = 1$ (black squares), $0.5$ (red triangles), and $0.3$ (blue circles). 
\label{fig7}}
\end{figure}

\subsection{Energy balance and effects of internal absorption}

The above estimate on the optical depth at the line wavelengths suggests that there might be non-negligible extinction operating on the intrinsic spectrum, and thus the dust properties estimated in \S 3.2 may need some modifications. In Figure 7, we show the `intrinsic' NIR spectrum corrected for dust absorption. For the dust obscuration, we assumed that the optical depth has the wavelength dependence of $\tau_{\lambda} \propto \lambda^{-1.4}$ and the normalization is taken at the optical depth derived at the wavelength of P$_{\beta}$. This relation represents very well the carbon dust properties in the wavelength range of interest with the typical size of $a_{\rm d} \lsim 0.01 \micron$ (see Fig. 6). The dust thermal emission luminosity needed to fit the observed spectrum is $\sim 5.1 \times 10^{42}$ erg s$^{-1}$, and it increases to $\sim 6.6 \times 10^{42}$ erg s$^{-1}$ in fitting the `intrinsic' spectrum. 

Fitting this `intrinsic' spectrum by thermal emission from carbon dust grains, as we did for the observed spectrum, we derive $(T_{\rm d}, M_{\rm d}) \sim (1450 {\rm K}, 7.5 \times 10^{-4} M_{\odot})$. Naively one would expect that the required dust mass would go up if one includes the dust absorption, but we find that it is indeed not necessarily the case. Namely, it has been frequently mentioned that fitting an observed spectrum with the dust emission under the `fully optically thin' assumption provides a lower limit of the dust mass as there should be additional dust particles hidden in the opaque region, but from our analysis we conclude that it can operate in the opposite way. With absorption, the required dust temperature and thus the specific intensity go up, and this effect reduces the dust mass needed to explain a given thermal emission flux. These two effects (color change and obscuration) almost cancel one another for SN 2010jl, and the resulting dust mass is indeed not very different in the two situations. Thus, we conclude that we can pin down the properties of the newly formed dust particles into a narrow range, namely $T_{\rm d} = 1350 - 1450$K and $M_{\rm d} = (7.5 - 8.5) \times 10^{-4} M_{\odot}$. In reality, the emission and absorption are coupled, and the transfer effects could make for an inhomogeneous temperature structure. Thus, radiation transfer simulations should be performed to be more conclusive on these properties, but this is beyond the scope of the present paper. Indeed, the optical depth we derived for the NIR wavelengths is below unity, thus the transfer effect should not be large anyway.

The same exercise can be performed for the optical spectrum. The observed luminosity in the optical wavelength is $\sim 1.4 \times 10^{42}$ erg s$^{-1}$ (obtained by integrating the HCT spectrum in Figure 1). Assuming the wavelength-dependent optical depth as described above, we obtain the intrinsic optical luminosity as $\sim 7.7 \times 10^{42}$ erg s$^{-1}$. Namely, the luminosity absorbed by the dust particles (in the optical wavelength) is $\sim 6.4 \times 10^{42}$ erg s$^{-1}$. It is comparable to the NIR thermal emission luminosity. However, there can be additional absorption occurring in the UV range which is not considered here. This could thus be an overestimate of the absorption in optical, and given the uncertainty in our treatment, we could allow a small difference in the optical depth for the lines and continuum. Let us now assume $\tau_{\rm cont} (\lambda) = \alpha \tau_{\rm line} (\lambda)$, where $\tau_{\rm cont}$ is the optical depth for the optical continuum emission while $\tau_{\rm line}$ is the optical depth derived from the line profile argument. In Figure 7, we show three examples, where we adopt $\alpha = 1$, 0.5, and 0.3. The case with $\alpha = 1$ corresponds to the case as described above. For $\alpha = 0.5$ and $0.3$, we obtain an intrinsic optical luminosity of $3.1$ and $2.2 \times 10^{42}$ erg s$^{-1}$, respectively. The moderate case with $\alpha \sim 0.5$ is consistent with the NIR luminosity with additional absorption in the UV. We note that the value of $\alpha$ smaller than unity would be consistent with a possible contribution of a CSM echo at a larger distance (see Appendix), since only a negligible fraction of the echo light would be absorbed by the newly formed dust grains. 

\begin{figure}
\hspace{-3cm}
        \begin{minipage}[]{0.7\textwidth}
                \epsscale{0.85}
                \plotone{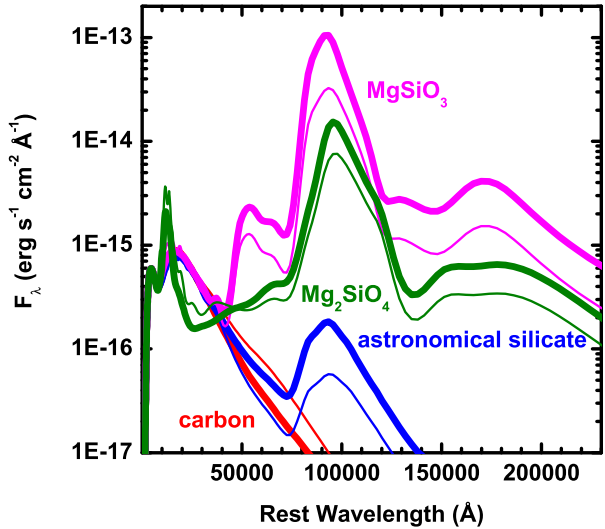}
        \end{minipage}
\vspace{-0.5cm}
\caption
{Expected dust thermal emission at IR wavelengths (up to $23\micron$), for carbon (red), astronomical silicate (blue), MgSiO$_3$ (magenta) and Mg$_2$SiO$_4$ (green). Those with $a_{\rm d} = 0.001 \micron$ (thick lines) and $1 \micron$ (thin lines) are shown. The macroscopic parameters, i.e., $T_{\rm d}$ and $M_{\rm d}$, are the same as shown in Figure 4. 
\label{fig8}}
\end{figure}

The above argument additionally shows that the carbon dust grains provide a consistent view here. Indeed, the balance between the absorption in optical and the emission in IR provides an additional clue to distinguish different dust species.  Figure 8 shows the expected thermal emission from different dust species in mid-IR, with the same model parameters ($T_{\rm d}$, $M_{\rm d}$) as in Figure 4 (which gives a good fit to the observed NIR spectrum for astronomical silicate and MgSiO$_3$, and a rough fit to the NIR flux for Mg$_2$SiO$_4$). Silicate dust, especially those without Fe, predicts a big bump at $\sim 10 \micron$. While no observations in such a wavelength range have been presented in this paper, it should affect the thermal balance. The resulting dust thermal emission luminosity is estimated to be $6.7 \times 10^{42}$ erg s$^{-1}$ (astronomical silicate), $6.3 \times 10^{44}$ erg s$^{-1}$ (MgSiO$_3$), and $1.1 \times 10^{44}$ erg s$^{-1}$ (Mg$_2$SiO$_4$). The astronomical silicate can still be consistent with the thermal balance (although carbon grains are preferred by other arguments), while MgSiO$_3$ and Mg$_2$SiO$_4$ are again both rejected by this argument since these dust particles should  emit mostly in mid-far IR (Fig. 8). The expected thermal emission would exceed the available energy budget if these were responsible for the NIR emission. 

The effect should be seen in the light curve as well. Here, adopting the properties of the dust grains obtained independently (\S 3.3), we estimate the effect of the newly formed dust in the optical light curve and compare the expected behavior with the observed light curve. Figure 7 also shows how much the observed photometry is changed due to dust absorption in the $B$, $V$, $R$, and $I$ bands (for carbon dust grains). SN 2010jl showed a flat light curve in the early phase. The light curve of SN 2010jl connecting the early \citep{zhang2012} and the late phases is shown in Figure 3. The decline rate in each band during $\sim 100 - 200$ days was fairly linear, and the rate per 100 days was $\sim 0.22$ mag in $B$, $\sim 0.27$ mag in $V$, $\sim 0.13$ mag in $R$, and $\sim 0.32$ mag in $I$ \citep{zhang2012}. The magnitude at $388$ days after the $V$-band maximum was already below the extrapolation from this decline rate \citep{zhang2012}, indicating that the dust formation and the subsequent absorption had already started at this epoch -- inspection of data presented by \citet{zhang2012} shows that the line profiles of H$_{\alpha}$ and H$_{\beta}$ showed a change (blueshift) around this epoch (\S 3.1), and the H$_{\alpha}$ luminosity evolution also changed its behavior around this epoch. Assuming that the `intrinsic' SN luminosity has evolved following the linear behavior observed in the earlier phase before the dust formation, we expect that the unobscured magnitudes were likely $B \sim 15.8$ mag, $V \sim 15.6$ mag, $R \sim 14.5$ mag, and $I \sim 15.2$ mag at $+553$ days. Thus, we estimate that the change in the magnitude due to the dust obscuration was $\sim 1.3$ mag in $B$, $\sim 1.2$ mag in $V$, $\sim 1.4$ mag in $R$, and $\sim 0.8$ mag in $I$ at $+553$ days, as is shown in Figure 3. If we assume the maximum possible contamination from the unresolved background (\S 2.2), this could reach to $\sim 2$ mag in $B$ and $V$ while $\sim 1.5$ mag in $R$ and $I$. The difference is roughly consistent with the dust obscuration with $\alpha$ in the range between 0.5 and 1. We also note that the estimated magnitude differences are generally larger for shorter wavelengths, consistent with the dust extinction scenario. Given our crude treatment, the value of $\alpha$ ($\sim 0.5 - 1$) indicates that the continuum and lines have experienced similar obscuration, suggesting that these two regions are roughly overlapping. 

Note that the above argument is essentially identical to the heating balance frequently considered in studying the dust properties \citep[see, e.g., ][]{nozawa2008}. Since we have confirmed that the absorbed luminosity in optical is consistent with the emission luminosity in NIR, with the dust temperature and the dust absorption coefficient we estimated independently, this means that the dust temperature we derived from the thermal emission is consistent with the expectation from the heating balance. 

The argument presented in this section shows that the optical depth of the newly formed dust grains is largely consistent with the observed light curve evolution without a significant change in the behavior of the intrinsic light curve evolution -- a frequently adopted assumption in discussing the dust formation. Indeed, this is a widely accepted (while practically difficult) test when the optical light curve shows accelerated fading to distinguish whether the decrease is due to the intrinsic change in the light curve evolution (i.e., strength of the CSM interaction, which could arise from a change in the CSM distribution) or mostly due to the absorption by the new source of absorption (i.e., newly formed dust grains). Generally it is difficult to distinguish the two, but in this paper we are able to do this thanks to an independent measurement of the optical depth from the dust particles. Our analysis indicates that the absorption effect is the main cause of the change in the optical luminosity decline rate, while we do not reject a minor contribution from a possible change in the strength of the interaction. 

\begin{figure}
\hspace{-2.5cm}
        \begin{minipage}[]{0.7\textwidth}
                \epsscale{0.8}
                \plotone{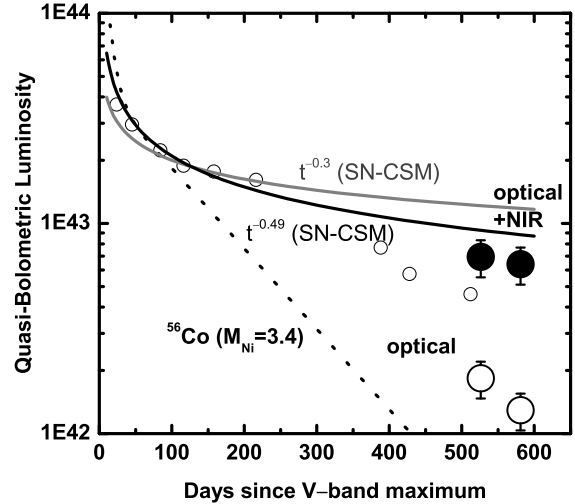}
        \end{minipage}
\vspace{-0.5cm}
\caption
{A roughly constructed, quasi-bolometric light curve of SN 2010jl (see \S 3.4 for details). The `optical' luminosity is shown by open circles \citep[where the small symbols are from ][]{zhang2012}. The bolometric luminosities from our observations are shown by the large circles, those including only the optical contribution (open circles) and those including both the optical and NIR (filled circles). The maximum possible contribution (i.e., the upper limit) of the contribution from the $^{56}$Co decay is shown by the dotted line, which is negligible at $\sim 550$ days. Power law behaviors expected from a smoothly distributed CSM interaction are shown by the solid curves at different power law indexes (where the power law index of $-0.3$ and $-0.49$ covers expectations from the CSM distribution expressed by a steady mass loss with a range of the ejecta structure) \citep[see, e.g., ][]{chevalier1982,moriya2013}. Given the rough treatment in constructing the bolometric luminosity, the `true' bolometric light curve (optical plus NIR) is fairly consistent with the simple `smooth' interaction model. 
\label{fig9}}
\end{figure}

The above point is highlighted in Figure 9, which shows a (roughly constructed) quasi-bolometric light curve of SN 2010jl, with and without the NIR contribution. Without observations in UV, we simply assume that the UV contributes $10\%$ of the optical luminosity following \citet{zhang2012} -- this is very likely an underestimate, given that the blue portion of the spectrum shows an excess as compared to a black body of $\sim 7,000$K. Given this uncertainty, we adopt a conservative error of $\pm 30$\% for the bolometric luminosity. Figure 9 also compares the observed light curve with a simple power law behavior which is expected from the CSM interaction with smooth CSM distribution. While the `optical' luminosity at $> 500$ days is well below the expectation from a smoothly-evolving CSM interaction, the `true' bolometric luminosity in this phase as a sum of the optical and NIR emissions (mostly NIR) is largely consistent with this expectation. Thus, we do not require a significant change in the strength of the interaction (or CSM distribution), and conclude that the obscuration by the newly formed dust is mainly responsible for the optical light curve evolution as well. There is a hint of slight decrease in the total luminosity as compared to the expectation, which might suggest that the strength of the CSM interaction might have changed slightly (but not at the level of being the main reason of the optical luminosity decrease). However, given our crude estimate of the bolometric luminosity, we regard this discrepancy as only indicative. We also note that the energy input from the decay of $^{56}$Co can provide at most only a minor contribution in the luminosity in the late phase (i.e., $> 100$ days), with the upper limit placed by the requirement that the decay luminosity cannot exceed the early phase luminosity ($< 100$ days) (note that we do not insist that the early phase was powered by the decay, but rather we simply place the upper limit). 

\subsection{Properties of the ejecta and CSM}

So far in this paper the dust emission and absorption have been treated independently, and an important question is if these two are mutually consistent, or what implications the combination of the two have. The emission line profiles suggest that the emitting region is extended to the velocity of $V \sim 4,000$ km s$^{-1}$, and it did not evolve much from the early to the late phase \citep{zhang2012}. The wavelength-dependence of the line profiles suggests that the line-emitting region and the dusty region are largely overlapping (\S 3.1 \& \S 3.3). Under the assumption that a dust mass of $M_{\rm d} \sim 8 \times 10^{-4} M_{\odot}$ is distributed in a shell whose inner radius is $R_{\rm d}$, the optical depth is estimated as follows: 
\begin{eqnarray}
& \tau_{\lambda} & \sim \frac{\kappa_{\lambda} M_{\rm d}}{4 \pi R_{\rm d}^2} \nonumber\\
 & \sim & 8 \left(\frac{\kappa_{\lambda}}{22,000 \ {\rm cm}^2 \ {\rm g}^{-1}}\right) 
\left(\frac{V}{4,000 \ {\rm km} \  {\rm s}^{-1}}\right)^{-2} 
\left(\frac{t}{550 \ {\rm days}}\right)^{-2} \ .
\end{eqnarray}
Inserting the opacity for carbon dust grains with $\lsim 0.1 \micron$, we obtain 
$\tau_{\lambda} \sim 10$ at 6563\AA\ (H$_{\alpha}$), $4$ at 12818\AA\ (P$_{\beta}$), and $2$ at 21661\AA\ (Br$_{\gamma}$). Namely, the opacity provided by the dust of $M_{\rm d}\sim 8 \times 10^{-4} M_{\odot}$ as has been obtained by the dust thermal {\em emission}  (\S 3.2) is by a factor of $\sim 10$ larger than that required to account for the dust {\em absorption} (\S 3.3). 

In other words, the dust inferred by the thermal emission will totally black out any SN emission if this would be distributed homogeneously in the SN emitting region. Thus, the dust formation must have been localized within the ejecta, with the `filling factor' being $\sim  0.1$, suggesting that the dust forming region is distributed as clumps within the ejecta. 

The dominance of the hydrogen emission lines suggests that the dust formation region is either the shocked H-rich CSM or shocked H-rich envelope of the SN ejecta. Assuming that all the carbon atoms with the mass fraction of $X(C)$ were condensed into the dust particles within the swept-up region, then the dust mass ($M_{\rm d}$) and swept-up mass ($M_{\rm sw}$) are connected as follows: $M_{\rm d} = M_{\rm sw} X(C)$. For the CSM case, by taking the mass fraction of carbon atoms normalized by the solar abundance, this provides the estimate on the swept-up mass as 
$M_{\rm sw} \sim 0.27 M_{\odot} (Z/Z_{\odot})^{-1}$, where $Z$ is the metallicity of the SN site. Since the condensation of atoms into the dust may not be complete, this provides a lower limit for $M_{\rm sw}$, namely $M_{\rm sw} \ge 0.27 M_{\odot} (Z/Z_{\odot})^{-1}$. The environmental metallicity of SN 2012jl has been reported to be low, $Z \lsim 0.3 Z_{\odot}$ \citep{stoll2011} (see also Appendix). Thus, we constrain the swept up mass as $M_{\rm sw} \gsim 0.9 M_{\odot}$. If this was ejected by a steady mass loss, it corresponds to a mass loss rate of $\gsim 0.02 M_{\odot}$ yr$^{-1}$ if the mass loss velocity was $\sim 100$ km s$^{-1}$ as was inferred from an  emission component in a high resolution spectrum \citep[or $6 \times 10^{-3} M_{\odot}$ yr$^{-1}$ with the wind velocity of $28$ km s$^{-1}$ as inferred from an absorption component: ][]{smith2011}. The average CSM density up to $\sim 2 \times 10^{16}$ cm is $n_{\rm CSM} \gsim 3 \times 10^{7}$ cm$^{-3}$, thus at least four or five orders of magnitude larger than typical CSM density around SNe Ib/c from a Wolf-Rayet progenitor \citep[e.g., ][]{chevalier2006b,maeda2013}, and at least two orders of magnitude larger than SNe IIp from a Red Supergiant progenitor \citep[e.g., ][]{chevalier2006a}.  We note the above estimate should roughly apply even if the dust formation site is the shocked H-rich SN ejecta, since the amount of the SN ejecta swept up by the reverse shock is comparable to the amount of the swept up CSM by the forward shock \citep{chevalier1982}, and since the metal content in the H-rich envelope should be similar to that in the CSM. The high-density CSM derived here may indicate LBV-like eruptions, suggested to take place at the end of the evolution of a massive star with $M_{\rm ms} \gsim 30 M_{\odot}$ \citep{smith2011}.

The above estimate assumes that the mass fraction of carbon atoms in the dust forming site is represented by the environmental (or progenitor) metallicity. However, this may well provide merely a conservative lower limit for the swept-up mass, if one considers possible effects of the stellar evolution on the carbon abundance in the progenitor surface and the CSM. If the CSM was indeed created by LBV eruptions, then the carbon abundance there is likely suppressed due to the CNO cycle as was observed in N-rich ejecta around Eta Carina \citep{davidson1986,smith2004}. This will lead to an increase in the estimate of $M_{\rm sw}$. On the other extreme, one may consider the case where the C-rich materials processed by the He-shell burning may have increased the carbon abundance in the progenitor surface and the CSM. We judge this latter case is highly unlikely, since such a case will lead to abundant He content in the interaction region as was observed for SN Ibc 2006jc \citep[e.g., ][]{anupama2009}, not the H-rich abundance inferred from the spectrum of SN IIn 2010jl. 

We can place a conservative upper limit on the mass loss as well (under the assumption that the carbon atom fraction is represented by $Z$). The `rarefied' region should be at a lower density than the dusty clumps. Taking into account the filling factor of about 0.1, then the mass loss properties cannot be larger than the above estimate by more than one order of magnitude. Thus, our constraints on the mass loss properties are the following: 

\noindent
{\bf The swept up mass:}\\ 
$0.9 (Z/0.3 Z_{\odot})^{-1} M_{\odot }\lsim M_{\rm SW} \lsim 9 (Z/0.3 Z_{\odot})^{-1} M_{\odot}$. \\
{\bf The mass loss rate (for a wind velocity of 100 km s$^{-1}$):}\\ 
$0.02 (Z/0.3 Z_{\odot})^{-1} M_{\odot}$ yr$^{-1}$ $\lsim \dot M \lsim 0.2 (Z/0.3 Z_{\odot})^{-1} M_{\odot}$ yr$^{-1}$. \\
{\bf The average density up to $\sim 2 \times 10^{16}$ cm:}\\ 
$3 \times 10^7 (Z/0.3 Z_{\odot})^{-1}$ cm$^{-3} \lsim n_{\rm CSM} \lsim 3 \times 10^8 (Z/0.3 Z_{\odot})^{-1}$ cm$^{-3}$. 

The CSM density we estimated here can be further checked with the relative strengths of the hydrogen emission lines. Assuming that the interaction region is largely ionized, we expect that the electron number density is comparable to $n_{\rm CSM}$, i.e., $n_{\rm e} \sim 10^8$ cm$^{-3}$. At this electron density, the H$_{\beta}/{\rm H}_{\alpha}$ ratio can be smaller than the Case B recombination if the Sobolev (gas) optical depth to H$_{\alpha}$ is large ($\tau_{\rm H_{\alpha}} \gsim 10$) \citep{xu1992}. If we fit the blue wing of H$_{\beta}$ to that of H$_{\alpha}$ (to estimate the Balmer line ratios in the unobscured wavelength), the intrinsic H$_{\beta}/{\rm H}_{\alpha}$ ratio of $\sim 0.1$ is derived, which is consistent with the expectation based on the derived CSM density ($n_{\rm CSM} \sim 10^8$ cm$^{-3}$), if $\tau_{{\rm H}_{\alpha}} \sim 10 - 100$ \citep{drake1980,xu1992}. The electron density cannot be larger than $\sim 10^{12}$ cm$^{-3}$, since then the H$_{\beta}/{\rm H}_{\alpha}$ ratio increases above the Case B recombination expectation \citep{drake1980,levesque2012}. Thus, we conclude that the CSM density we derived is consistent with the hydrogen line strengths. 

Now, with the properties of the CSM estimated above through the dust properties, we can check if the expected properties of the SN arising from the SN-CSM interaction are consistent with the observations. The energy budget ($E$) and energy generation rate at the shock ($L$) due to the interaction are \citep[see also, ][]{zhang2012}, 
\begin{eqnarray}
E &\sim& 8 \times 10^{50} \left(\frac{M_{\rm sw}}{5 M_{\odot}}\right) \left(\frac{V}{4,000 \ {\rm km} \ {\rm s}^{-1}}\right)^2 \ {\rm erg} \ , \\
L &\sim& 2 \times 10^{43} \left[\frac{\dot M / 0.1 M_{\odot} \ {\rm yr}^{-1}}{v_{\rm w} / 100 \ {\rm km} \ {\rm s}^{-1}}\right] \left(\frac{V}{4,000 \ {\rm km} \ {\rm s}^{-1}}\right)^3 \ {\rm erg} \  {\rm s}^{-1}
\end{eqnarray}
These are comparable to the observed radiation output ($\sim 4 \times 10^{50}$ erg in the first 200 days, which should be added by the smaller amount of the energy emitted after $+200$ days) and the total observed luminosity in optical and NIR at $+550$ days ($\sim 6.5 \times 10^{42}$ erg), assuming that about half of the kinetic energy was converted to the radiation energy. Detailed analysis is required to improve the above estimate, to address the conversion efficiency of the kinetic energy to the radiation energy and the shock velocity evolution \citep{moriya2012}, but the rough agreement above strongly supports that the SN-CSM interaction is the main power source of the luminous SN IIn 2010jl. 

\section{Discussion and Conclusions}

\subsection{Summary}

In this paper, we presented optical and NIR spectra of the luminous SN IIn, 2010jl, at $\sim + 550$ days since $V$-band maximum. The NIR spectrum clearly shows a thermal continuum emission, which was absent in the early phase. The hydrogen emission lines show a wavelength-dependent line profile, where the optical lines show large blueshifts, P$_{\beta}$ shows a small blueshift, and there is virtually no shift for Br$_{\gamma}$. With these two spectroscopic features, we conclude that the dust grains were newly formed in SN IIn 2010jl in the late phase. The flux in the optical bands at $\sim +550$ days is offset from an extrapolation of the flat evolution seen in the early phase up to $\sim +200$ days. This indicates that SN 2010jl showed additional evidence for dust formation, namely the optical light is absorbed and re-emitted at longer wavelengths. Thus, SN 2010jl in the late phase has all the expected properties of new dust formation. The dust we see in the late phase ($\sim 550$ days) was most likely formed around 1 year after the $V$-band maximum. 

We derived the mass and temperature of the newly formed dust grains. Thanks to the spectroscopic detection, the properties of the dust grains are specified with little uncertainty. 

\noindent 
{\bf The main species:}\\
Carbon grains, either amorphous carbon or graphite. \\
{\bf The typical size of the dust grains:}\\
$a_{\rm d} \lsim 0.1 \micron$ (most likely $a_{\rm d} \lsim 0.01 \micron$).\\
{\bf The dust temperature:}\\
$T_{\rm d} \sim 1,350 - 1,450$K.\\
{\bf The total mass of the dust grains}:\\
$M_{\rm d} \sim (7.5 - 8.5) \times 10^{-4} M_{\odot}$.

Note that the range of values of $T_{\rm d}$ and $M_{\rm d}$ takes into account the uncertainty in the optical depth of the dusty region to the NIR photons. The argument for the carbon grains is not only based on the NIR SED, but also on the opacity dependence with wavelength as derived by the different line profiles of hydrogen lines at different wavelengths. We have checked the heating balance between the absorption in optical and the emission in NIR, and we conclude that this is consistent with the properties of the dust (e.g., temperature) that we have derived by an independent method. Non-astronomical silicate grains like MgSiO$_3$ or Mg$_2$SiO$_4$ are clearly rejected by various arguments. 

With the properties of the dust grains, we estimated the properties of CSM and SN ejecta as follows: 

\noindent
{\bf The swept up mass:}\\ 
$0.9 (Z/0.3 Z_{\odot})^{-1} M_{\odot }\lsim M_{\rm SW} \lsim 9 (Z/0.3 Z_{\odot})^{-1} M_{\odot}$. \\
{\bf The mass loss rate (for the wind velocity of 100 km s$^{-1}$):}\\ 
$0.02 (Z/0.3 Z_{\odot})^{-1} M_{\odot}$ yr$^{-1}$ $\lsim \dot M \lsim 0.2 (Z/0.3 Z_{\odot})^{-1} M_{\odot}$ yr$^{-1}$. \\
{\bf The average density up to $\sim 2 \times 10^{16}$ cm:}\\ 
$3 \times 10^7 (Z/0.3 Z_{\odot})^{-1}$ cm$^{-3} \lsim n_{\rm CSM} \lsim 3 \times 10^8 (Z/0.3 Z_{\odot})^{-1}$ cm$^{-3}$. \\
{\bf The structure of the interacting region: }\\
The dust grains are localized within  $\sim 10$\% of the emitting volume. 

The CSM properties correspond to the mass loss history of the progenitor in the last 60 years before the explosion, assuming the mass loss wind velocity of $\sim 100$ km s$^{-1}$. The localized formation of the dust grains indicates that the dust grains were formed in high-density regions, probably distributed in high-density clumps. This supports the formation of the dust in a dense cooling shell created by the interaction between the SN and a dense CSM, as the dense cooling shell (or clumps) is expected to be formed from regions which experience radiative and hydrodynamic instabilities. The high density CSM is further supported by the relative strength of hydrogen emission lines. 

We thus conclude that there is a very dense CSM around SN 2010jl, and we have found that the interaction of the SN with the CSM defined above can roughly explain the luminosity and energy output of the luminous SN 2010jl. The high-density CSM was likely produced by LBV-like eruptions, suggested to take place at the end of the evolution of a massive star with $M_{\rm ms} \gsim 30 M_{\odot}$ \citep{smith2011}.

\subsection{Discussions}

As shown in this paper, the dust formation scenario provides a consistent and straightforward interpretation to various features of SN 2010jl in the late phase. Still, one may consider different scenarios, which may, in principle, partly contribute to each feature to some extent. Indeed, a combination of the blueshifted H${_\alpha}$ profile evolution and the flat optical light curve evolution up to $\sim 400$ days led \citet{zhang2012} to hypothesize that these could be explained by the CSM interaction without invoking the dust formation in this early epoch (before the epoch of the dust formation we propose in this paper). If the extension of this idea could explain the later evolution, i.e., further blueshift in the H$_{\alpha}$ emission and the accelerated fading, is another story. For example, one may argue that the accelerated fading could be caused by the end of the main CSM interaction that could be related to the possible decrease in the $H_{\alpha}$ blueshift at $\gsim 500$ days as compared to that at $\sim 400$ days (while at $\gsim 500$ days it was still larger than $\sim 200$ days) \citep{zhang2012}. However, since the starting epoch of the change in this optical luminosity decrease and that of the possible $H_{\alpha}$ blueshift decrease were not coincident, we speculate that such a scenario would require a rather fine tuning. In the dust formation scenario, the decrease in the degree of blueshift seen in $H_{\alpha}$ at $\gsim 500$ days could be naturally explained if the dust formation was more or less completed (which we attributed to be around 1 year since the $V$-band maximum) and then the optical depth simply decreased following the expansion. In any case, it is definitely interesting to investigate details of the light curve and the line profile evolution, and their mutual relation (which has not been discussed quantitatively so far). Such a study will lead to an accurate evaluation on the properties of the dust grains. 

Investigating details of the CSM interaction is however beyond the scope of the present paper, but we present some arguments that this would not change our result significantly. For example, this would not cause the wavelength-dependent line shift. Given that the optical depth we estimate based only on the line profiles at various hydrogen emission lines is quite consistent with the difference between the observed optical luminosity at $\sim 550$ days and the extrapolation from the early phase light curve evolution, there is no need to introduce additional cause of the luminosity change. Furthermore, we showed that the bolometric luminosity including NIR does not show significant change in the decline rate, strongly indicating that the possible effect from the change in the CSM interaction property should provide at most only a minor contribution. 

We estimate that the dust particles we observed were formed at $\sim 1$ year after the explosion. There were possible dust signatures reported based on observations of SN 2010jl in the earlier phase, and we conclude that they are not the same component as described in this paper. \citet{andrews2011} detected IR excess at $3.6$ and $4.5 \micron$ (with the peak redder than $4.5 \micron$), attributing it to an echo by pre-existing CSM dust grains. The NIR dust emission in the late phase exceeds the NIR magnitudes at day 108 reported in \citet{andrews2011}.With the monotonically decreasing SN luminosity \citep{zhang2012}, it is unlikely that the dust we observed was this pre-existing CSM dust heated by the SN emission. Also, the estimated position of the CSM dust ($\sim 10^{18}$ cm from the progenitor) is too far to be shock heated by the SN forward shock.  Indeed, if we estimate the location of the materials emitting the NIR thermal continuum assuming a black body (while this is just a rough estimate, since the optical depth in NIR as we derived is below unity), we obtain a radius of $\sim 2.8 \times 10^{16}$ cm from our NIR spectrum. This is much smaller than the pre-existing dust found by \citet{andrews2011}. The black body estimate is indeed close to the expected radius reached by the shock wave ($\sim 2 \times 10^{16}$ cm for a constant shock velocity of $4,000$ km s$^{-1}$, which is very likely an underestimate since the shock must have experienced deceleration), strengthening the case for the newly formed dust grains at the dense-cooling shell. 
Thus, the dust particles we observed in the late phase cannot be the same (pre-existing) dust particles reported in the early phase. Indeed, the above estimate of the emitting radius ($\sim 2 - 3 \times 10^{16}$ cm) also generally excludes a contribution from any pre-existing dust grains to the NIR emission at $\sim 550$ days, since at such a small radius and with a large luminosity of SN 2010jl, any pre-existing dust grains there must have been destroyed by the SN radiation \citep[e.g., ][]{fox2010}. 

In this context, we emphasize that our argument in the heating-cooling balance (\S 3.4) is totally independent from any pre-existing CSM dust grains. Such pre-existing dust grains never produce a `change' in the absorption (thus the decrease in the optical luminosity). The possible mid-IR emission from the pre-existing dust must not be added to the bolometric luminosity from the system, since these dust grains at the distance of $\sim 10^{18}$ cm \citep{andrews2011} would reprocess the early phase radiation (thus should not be counted in considering the instantaneous luminosity output). From this argument, we conclude that we cannot either support or reject the pre-existing CSM dust from our observations, but our analysis of the newly formed dust grains is essentially insensitive to this. 

\citet{smith2012} reported wavelength dependent blueshift in the hydrogen lines in the early phase ($\lsim +100$ days), and they attributed this behavior either to the newly formed dust or to the optical depth effect of the emitting gas. We note that the blueshift we observe is more significant, as is shown by the H$_{\alpha}$ line profile evolution from the early to late phase \citep{zhang2012}. Although no detailed analysis on the dust properties (including the mass and temperature) was presented in \citet{smith2012}, we suspect that the dust mass must have been very small if not zero -- otherwise the wavelength shift should have been greater in the early epoch and the NIR flux should have been much larger than in the late phase. We speculate that already in the early phase a small fraction of the shocked ejecta/CSM  might have started the dust formation, but the major dust formation took place at $\sim 1$ year after the explosion. 

We conclude that the newly formed dust grains are carbon grains. One interesting question is how the carbon dust grains were formed. As we conclude that the dust formation took place either in the shocked H-rich CSM or the shocked H-rich SN envelope, we expect that the number fraction of oxygen atoms was higher than that of carbon. In such a situation, the formation of carbon dust grains may be forbidden by the consumption of carbon atoms into the CO molecules \citep{nozawa2011}. Observationally, we have not detected any clear signature of CO molecules in the $K$-band spectrum despite good S/N -- thus, it seems that even within the O-rich gas the carbon dust grains can form by avoiding the formation of CO molecules. The investigation of such a process is beyond the scope of this paper \cite[see, e.g.,][]{clayton2013}, but we note that this could provide a hint for understanding the dust formation in SNe and other astrophysical environments. It would also be interesting to see if the carbon grain formation would have been preceded by formation of CO molecules.  Unfortunately, the Spitzer data around 100 days \citep{andrews2011}, well before our claimed epoch for the major formation of the dust grains, do not either support or reject existence of the CO. This anyway highlights the importance of both multi-epoch and multi-band observations. Another interesting possibility for the future is to follow SN 2010jl further to see if silicate dust is also formed as the temperature decreases. 

In addition to the dust properties, our analysis allows us to constrain the CSM environment, independent from the optical light curve. We note that the density and distribution have also been  constrained by X-ray observations. \citet{chandra2012} \citep[see also ][]{ofek2013} derived  the density slope of $\rho_{\rm CSM} \propto r^{-1.6}$ and the mass loss rate of $\sim 5 \times 10^{-3} M_{\odot}$ yr$^{-1}$ normalized at $10^{15}$ cm (here we assume the shock velocity of $4,000$ km s$^{-1}$ and the wind velocity of $100$ km s$^{-1}$ for a comparison to our result). This corresponds to the corresponding `local' mass loss rate of $\sim 0.017 M_{\odot}$ yr$^{-1}$ at the distance of $2 \times 10^{16}$ cm. Thus, our independent estimate of the mass loss rate through the dust formation signal is roughly consistent with the X-ray measurement. Given uncertainties in both methods (e.g., ionization status in analyzing the X-ray absorption), we regard that the agreement is fairly good. 

The conclusion about the high density CSM likely applies to other luminous SNe IIn as well. We note a striking similarity in the ejecta and CSM properties we derived for SN 2010jl and those derived for SN IIn 2006tf by \citet{smith2008b}, despite totally different approaches to estimate these properties. SN IIn 2006tf was similar to SN 2010jl in the luminosity, light curves, and spectral evolution \citep{zhang2012}. \citet{smith2008b} estimated the mass loss rate in the last decade before the explosion of SN 2006tf reached a few $M_{\odot}$ yr$^{-1}$ to explain the luminosity of SN 2006tf (similar to that of SN 2010jl) by the SN-CSM interaction. The CSM density we estimated is a bit lower than this, but still much higher than a usual stellar wind and close to the amount expected for LBV-like eruptions of a massive star. Given uncertainties in both estimates (i.e., from the interaction luminosity and from the dust mass), the situations we infer for these two SNe are strikingly similar. Our analysis on the CSM properties around SN 2010jl will provide a useful basis on analyzing other (luminous) SNe IIn and potentially (at least a part of) SLSNe. 

\acknowledgements 
This research is based on observations obtained at the Subaru Telescope (S12A-047) operated by the Astronomical Observatory of Japan, and at the Himalayan Chandra Telescope (HCT) operated by the Indian Institute of Astrophysics. The authors thank the staff at the Subaru Telescope and the Himalayan Chandra Telescope for their excellent support in the observations. The authors also thank Masayuki Tanaka, Filomena Bufano and Giuliano Pignata for valuable discussion. This research is supported by World Premier International Research Center Initiative (WPI Initiative), MEXT, Japan. This work has also been supported in part by the DST-JSPS Bilateral Joint Research Projects. K.M. and R.Q. acknowledge financial supports by Grant-in-Aid for Scientific Research for Young Scientists (23740141, 24740118). T.J.M. is supported by the Japan Society for the Promotion of Science Research Fellowship for Young Scientists (23-5929). We have used the Weizmann interactive supernova data repository (www.weizmann.ac.il/astrophysics/wiserep) to obtain the archive spectrum data for SN 2010jl around the $V$-band maximum. 

\appendix

\section{A. Contamination from the unresolved background region}

\begin{figure*}
\begin{center}
        \begin{minipage}[]{0.95\textwidth}
                \epsscale{0.7}
                \plotone{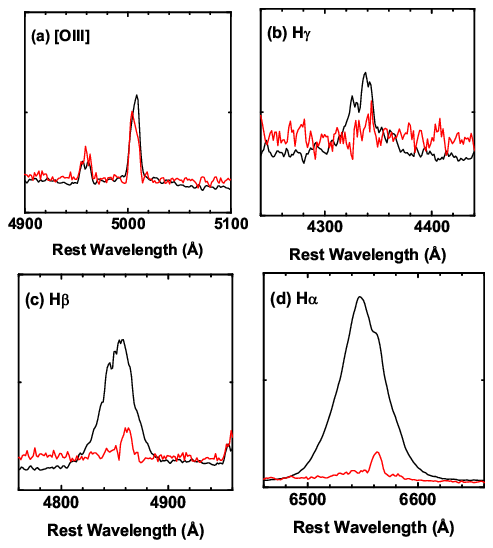}
        \end{minipage}
\end{center}
\vspace{-1cm}
\caption
{The comparison between the extracted SN spectrum (black) and the background spectrum (red). The background spectrum is scaled to the [OIII]~$\lambda\lambda$4959,5007 seen in the SN spectrum (a), as this is very likely an residual emission from the unresolved background region. The scaled background spectrum is then compared to the SN spectrum in the wavelength ranges for H$_{\gamma}$ (b), H$_{\beta}$ (c), and H$_{\alpha}$ (d). 
\label{figa1}}
\end{figure*}

In this section, we check how much the narrow component of the hydrogen emission lines is contaminated by emissions from the unresolved host and/or the SN environment, rather than a narrow emission from the CSM just around the SN.  We assume that the HCT spectrum of the SN vicinity, used for the background subtraction, represents a spectrum of the unresolved background region from the host galaxy and/or SN environment (not affected by the SN emission). We estimate a possible contribution from the unresolved background region to the narrow component of the hydrogen emission lines seen in the SN spectrum as follows (Fig. A1). The extracted SN spectrum (Fig. 1) shows narrow emission lines of [OIII]~$\lambda\lambda$4959, 5007, likely originated from the background, rather than the SN/CSM itself. We first scaled the background spectrum by the flux of the [OIII], then assumed that the scaled spectrum is representative of the unresolved background region. Then, we compared this spectrum with the hydrogen emission lines. As shown in Figure A1, the possible contamination from the background to the narrow emission lines seen in the SN spectrum is not negligible. Therefore, we decided not to discuss details of the narrow component in the main text. 

\section{B. SN site metallicity}

\citet{stoll2011} reported that the metallicity around SN 2010jl is low, $Z \lsim 0.3 Z_{\odot}$. With the SN vicinity spectrum we obtained, we cross-checked this estimate. We first measured the fluxes of the narrow emission lines, H$_{\alpha}$, H$_{\beta}$, [NII]~$\lambda$6584, and [OIII]~$\lambda$5007. Then we used the N2 indicator (${\rm N2} \equiv \log({\rm [NII]}/{\rm H}_{\alpha})$) and O3N2 indicator (${\rm O3N2} \equiv \log (({\rm [OIII]}/{\rm H}_{\beta})/({\rm [NII]}/{\rm H}_{\alpha}))$), in order to derive the oxygen abundance ($\log ({\rm O}/{\rm H})$) using relations in \citet{pettini2004}. We then derived the metallicity as $12 + \log({\rm O}/{\rm H}) \sim 8.31$ by the N2 diagnostics and $\sim 8.36$ by the O3N2 diagnostics. Given the typical uncertainty of $\sim 0.1 - 0.2$ dex in these methods, our result agrees with that presented by \citet{stoll2011}. Adopting the solar metallicity as $12 + \log({\rm O}/{\rm H}) = 8.66$ \citep{asplund2004} and representing the metallicity by the O abundance, we estimate that $Z \sim 0.3 - 0.7 Z_{\odot}$. While the mean is larger than that derived by \citet{stoll2011}, the two results still agree within the error. In the main text, we adopt $Z = 0.3 Z_{\odot}$ as a reference value for the metallicity, following \citet{stoll2011}.

\section{C. Possible echo contribution in the optical spectrum}

As mentioned in \S 3.2, our optical spectrum in the late phase ($\sim +550$ days) shows deviation from a single black body fit especially in the blue, either due to unresolved lines or a light echo caused by the CSM. While it is not a main focus of the present paper, this itself is an interesting question, thus we provide a constraint with our data on this issue in this Appendix. 

\begin{figure*}
\begin{center}
        \begin{minipage}[]{0.95\textwidth}
                \epsscale{0.7}
                \plotone{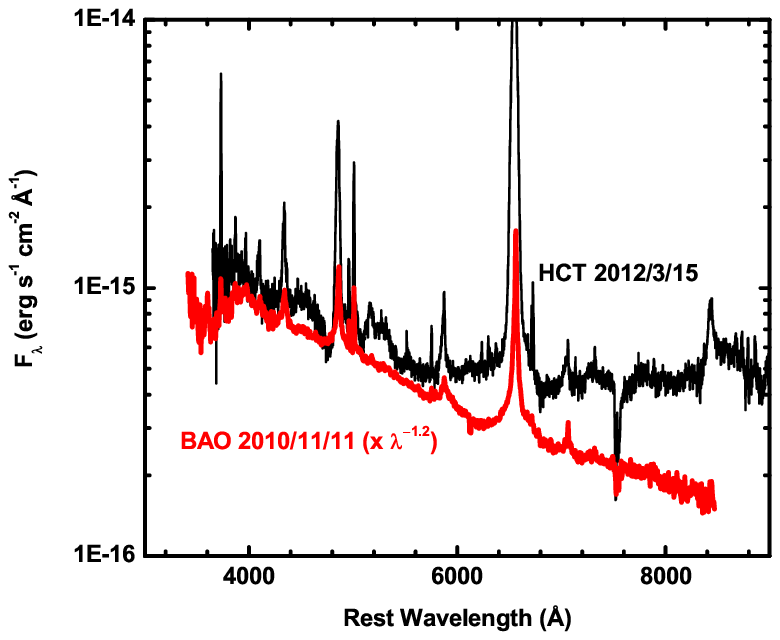}
        \end{minipage}
\end{center}
\vspace{-1cm}
\caption
{The possible contribution of the CSM echo in the late phase optical spectrum. The HCT spectrum (black) is compared with a pseudo echo spectrum (red), which was constructed from the spectrum around $V$-band maximum \citep{zhang2012}, multiplied by the further dependence of $f_{\lambda} \propto \lambda^{-1.2}$ to mimic the scattered spectrum. The flux in the `echo' spectrum is scaled as large as possible, while not to exceed the flux in the late phase spectrum. 
\label{fig21}}
\end{figure*}

Figure C1 shows a comparison between the late phase spectrum and a pseudo echo spectrum. For the pseudo echo spectrum, we simply adopt the maximum-phase spectrum at $\sim 24$ days since the $V$-band maximum \citep[][]{zhang2012}\footnote[15]{The spectrum was obtained through The Weizmann interactive supernova data repository (www.weizmann.ac.il/astrophysics/wiserep) \citep[][]{yaron2012}} and apply an additional dependence of $f_{\lambda} \propto \lambda^{-1.2}$ to mimic the scattered spectrum \citep{miller2010}. From this exercise, we see that there is a hint of a light echo contribution below $\sim 5,000$\AA, where the overall continuum slope is consistent with the expectation from the echo scenario. While this is not a probe of an echo contribution, this provides an upper limit of the contribution of an echo in the optical emission, which we estimate to be $\lsim 50$\% at maximum in luminosity. Since the thermal balance argument in \S 3.4 already includes the uncertainty at this level (e.g., in the value of $\alpha$), this would not change our conclusion. 

At the same time, there are emission features which cannot be attributed to the echo (e.g., at $\sim 4,500$ and $5,200$\AA). Thus, even if the echo is present, the unresolved lines are also present as suggested by, e.g., \citet[][]{foley2007}. Further investigating the echo contribution in the late phase emission is definitely interesting, and we will present detailed analysis on this issue elsewhere.


\begin{thebibliography}{}

\bibitem[Andrews et al. (2010)]{andrews2010}
Andrews, J. E., et al. 2010, ApJ, 715, 541

\bibitem[Andrews et al.(2011a)]{andrews2011}
Andrews, J.E., et al. 2011, AJ, 142, 45

\bibitem[Andrews et al.(2011b)]{andrews2011b}
Andrews, J.E., et al. 2011, ApJ, 731, 47

\bibitem[Anupama et al.(2009)]{anupama2009}
Anupama, G.C., Sahu, D.K., Gurugubelli, U.K., Prabhu, T.P., Tominaga, N., Tanaka, M., Nomoto, K. 
2009, MNRAS, 392, 894 

\bibitem[Aspulund et al.(2004)]{asplund2004}
Asplund, M., Grevesse, N., Sauval, A.J., Allende Prieto, C., \& Kiselman, D. 
2004, A\&A, 417, 751

\bibitem[Benetti et al.(2010)]{benetti2010}
Benetti, S., et al. 2010, CBET, 2536, 1

\bibitem[Chandra et al.(2012)]{chandra2012}
Chandra, P., Chevalier, R.A., Irwin, C.M., Chugai, N., Fransson, C., Soderberg, A.M. 
2012, ApJ, 750, L2

\bibitem[Chatzopoulos, Wheeler, \& Vinko(2012)]{chatzopoulos2012} 
Chatzopoulos, E., Wheeler, J.~C., Vinko, J., 2012, ApJ, 746, 121

\bibitem[Chevalier(1982)]{chevalier1982}
Chevalier, R.A., 1982, ApJ, 258, 790 

\bibitem[Chevalier \& Fransson(1994)]{chevalier1994}
Chevalier, R.A., \& Fransson, C. 1994, ApJ, 420, 268

\bibitem[Chevalier et al.(2006a)]{chevalier2006a}
Chevalier, R.A., Fransson, C., \& Nymark, T.K. 2006a, ApJ, 641, 1029

\bibitem[Chevalier \& Fransson(2006b)]{chevalier2006b}
Chevalier, R.A., \& Fransson, C. 2006b, ApJ, 651, 381 

\bibitem[Chevalier \& Irwin(2011)]{chevalier2011} 
Chevalier, R.~A., Irwin, C.~M., 2011, ApJ, 729, L6 

\bibitem[Chugai(2001)]{chugai2001}
Chugai, N.N. 2001, MNRAS, 326, 1448

\bibitem[Chugai et al.(2004)]{chugai2004}
Chugai, N.N.,et al. 2004, MNRAS, 352, 1213

\bibitem[Clayton(2013)]{clayton2013}
Clayton, D.D. 2013, ApJ, 762, 5

\bibitem[Davidson et al.(1986)]{davidson1986}
Davidson, K., Dufour, R.J., Walborn, N.R., Gull, T.R. 
1986, ApJ, 305, 867

\bibitem[Dorschner et al.(1995)]{dorschner1995}
Dorschner, J., Begemann, B., Henning, Th., Jaeger, C., Mutschke, H.
1995, A\&A, 300, 503

\bibitem[Draine(2003)]{draine2003}
Draine, B. T. 2003, ApJ, 598, 1026 

\bibitem[Drake \& Ulrich(1980)]{drake1980}
Drake, S.A., \& Ulrich, R.K. 1980, ApJS, 42, 351

\bibitem[Dwek et al.(2007)]{dwek2007}
Dwek, E., Galliano, F., \& Jones, A.P. 2007, ApJ, 662, 927

\bibitem[Filippenko(1997)]{filippenko1997}
Filippenko, A.V. 1997, ARA\&A, 35, 309

\bibitem[Foley et al.(2007)]{foley2007}
Foley, R.J., Smith, N., Ganeshalingam, M., Li, W., Chornock, R., Filippenko, A.V. 
2007, ApJ, 657, L105

\bibitem[Fox et al.(2009)]{fox2009}
Fox, O., et al. 2009, ApJ, 691, 650 

\bibitem[Fox et al.(2010)]{fox2010}
Fox, O.D., Chevalier, R.A., Dewek, E., Skrutskie, M.F., Sugerman, B.E.K., Leisenring, J. 
2010, ApJ, 725, 1768 

\bibitem[Fox et al.(2011)]{fox2011}
Fox, O.D., et al. 2011, ApJ, 741, 7

\bibitem[Gal-Yam(2012)]{galyam2012}
Gal-Yam, A. 2012, Science, 337, 927

\bibitem[Gerardy et al.(2000)]{gerardy2000}
Gerardy, C.L., Fesen, R.A., H\"oflich, P, Wheeler, J.C. 2000, ApJ, 119, 2968

\bibitem[Goto et al.(2003)]{goto2003}
Goto, M., et al. 2003, SPIE proc., 4839, 1117

\bibitem[Hayano et al.(2008)]{hayano2008}
Hayano, Y., et al. 2008, Proc. SPIE, 2008, 7015, 25

\bibitem[Hayano et al.(2010)]{hayano2010}
Hayano, Y., et al. 2010, Proc. SPIE, 2010, 7736, 21

\bibitem[Hawarden et al.(2001)]{hawarden2001}
Hawarden, T.G., Leggett, S.K., Letawsky, M.B., Ballantyne, D.R., Casali, M.M. 
2001, MNRAS, 325, 563

\bibitem[Horne(1986)]{horne1986}
Horne, K., et al. 1986, PASP, 98, 609 

\bibitem[Kobayashi et al.(2000)]{kobayashi2000}
Kobayashi, N., et al. 2000, Proc. SPIE, 4008, 1056

\bibitem[Kotak et al.(2009)]{kotak2009}
Kotak, R., et al. 2009, ApJ, 704, 306

\bibitem[Kozasa et al.(1989)]{kozasa1989}
Kozasa, T., Hasegawa, H., \& Nomoto, K. 1989, ApJ, 344, 325

\bibitem[Kozasa et al.(2009)]{kozasa2009}
Kozasa, T., Nozawa, T., Tominaga, N., Umeda, H., Maeda, K., Nomoto, K. 
2009, ASP conf. series, 414, 43

\bibitem[Levesque et al.(2012)]{levesque2012}
Levesque, E.M., Stringfellow, G.S., Ginsburg, A.G., Bally, J., Keeney, B.A. 
2012, ApJ, submitted (arXiv:1211.4577)

\bibitem[Lucy et al.(1989)]{lucy1989}
Lucy, L. B., Danziger, I. J., Gouiffes, C., \& Bouchet, P. 
1989, in IAU Colloq. 120, Structure and Dynamics of the Interstellar 
Medium, ed.G. Tenorio-Tagle, M. Moles, \& J. Melnick 
(LNP 350; Berlin: Springer),164

\bibitem[Maeda(2013)]{maeda2013}
Maeda, K. 2013, ApJ, 762, 14

\bibitem[Mattila et al.(2008)]{mattila2008}
Mattila, S., et al. 2008, MNRAS, 389, 141

\bibitem[Meikle et al.(1993)]{meikle1993}
Meikle,W. P. S., Spyromilio, J., Allen, D. A., Varani, G.-F., \& Cumming, R. J. 
1993, MNRAS, 261, 535 

\bibitem[Meikle et al.(2007)]{meikle2007}
Meikle, W.P.S., et al. 2007, ApJ, 665, 608 

\bibitem[Miller et al.(2010)]{miller2010}
Miller, A.A., Smith, N., Li, W., Bloom, J.S., Chornock, R., Filippenko, A.V., Prochaska , J.X. 
2010, ApJ, 139, 2218

\bibitem[Moriya et al.(2013a)]{moriya2012}
Moriya, T.J., Blinnikov, S.I., Tominaga, N., Yoshida, N., Tanaka, M., Maeda, K., Nomoto, K. 
2013a, MNRAS, 428, 1020

\bibitem[Moriya et al.(2013b)]{moriya2013}
Moriya, T.J., Maeda, K., Taddia, F., Sollerman, J., Blinnikov, SI., \& Sorokina, E.I. 
2013b, MNRAS, submitted

\bibitem[Moseley et al.(1989)]{moseley1989}
Moseley, S. H., Dwek, E., Glaccum, W., Graham, J. R., Loewenstein, R. F., \& Silverberg, R. F. 1989, Nature, 340, 697 

\bibitem[Newton \& Puckett(2010)]{newton2010}
Newton, J., \& Puckett, T. 2010, CBET, 2532, 1

\bibitem[Nozawa et al.(2003)]{nozawa2003}
Nozawa, T., Kozasa, T., Umeda, H., Maeda, K., \& Nomoto, K. 2003, ApJ, 598,
785

\bibitem[Nozawa et al.(2008)]{nozawa2008}
Nozawa, T., et al. 2008, ApJ, 684, 1343 

\bibitem[Nozawa et al.(2011)]{nozawa2011}
Nozawa, T., Maeda, K., Kozasa, T., Tanaka, M., Nomoto, K., Umeda, H. 
2011, ApJ, 736, 45

\bibitem[Ofek et al.(2013)]{ofek2013}
Ofek, E.O., et al. 2013, ApJ, 763, 42

\bibitem[Osterbrock (1989)]{osterbrock}
Osterbrock, D.E. 1989, Astrophysics of Gaseous Nebulae and Active Galactic Nuclei, University Science Books

\bibitem[Pastorello et al.(2008)]{pastorello2008}
Pastorello, A., et al. 2008, MNRAS, 389, 113

\bibitem[Pettini \& Pagel(2004)]{pettini2004}
Pettini, M., \& Pagel, B.E.J. 2004, MNRAS, 348, L59

\bibitem[Pozzo et al.(2004)]{pozzo2004}
Pozzo, M., et al. 2004, MNRAS, 352, 457

\bibitem[Quimby et al.(2011)]{quimby2011}
Quimby, R.M., et al. 2011, Nature, 474, 487

\bibitem[Sakon et al.(2009)]{sakon2009}
Sakon, I., et al. 2009, ApJ, 692, 546 

\bibitem[Schlegel et al.(1998)]{schlegel1998}
Schlegel, D.J., Finkbeiner, D.P., Davis, M. 
1998, ApJ, 500, 525

\bibitem[Semenov et al.(2003)]{semenov2003}
Semenov, D., Henning, Th., Helling, Ch., Ilgner, M., Sedlmayr, E.
2003, A\&A, 410, 611

\bibitem[Smith \& Morse(2004)]{smith2004}
Smith, N., Morse, J.A. 
2004, ApJ, 605, 854

\bibitem[Smith et al.(2008a)]{smith2008a}
Smith, N., Foley, R. J., \& Filippenko, A. V. 2008a, ApJ, 680, 568

\bibitem[Smith et al.(2008b)]{smith2008b} 
Smith, N., et al. 2008b, ApJ, 686, 467

\bibitem[Smith et al.(2011)]{smith2011}
Smith, N., et al. 2011, ApJ, 732, 63

\bibitem[Smith et al.(2012)]{smith2012}
Smith, N., et al. 2012, AJ, 143, 17

\bibitem[Stoll et al.(2011)]{stoll2011}
Stoll, R., et al. 2011, ApJ, 730, 34

\bibitem[Stritzinger et al.(2012)]{stritzinger2012}
Stritzinger, M., et al. 2012, ApJ, 756, 173

\bibitem[Sugerman et al.(2006)]{sugerman2006}
Sugerman, B.E.K., et al. 2006, Science, 313, 196

\bibitem[Suntzef \& Bouchet(1990)]{suntzeff1990}
Suntzeff, N. B., \& Bouchet, P. 1990, AJ, 99, 650

\bibitem[Taddia et al.(2013)]{taddia2013}
Taddia, F., et al. 2013, A\&A, in press (arXiv:1304.3038)

\bibitem[Todini \& Ferra(2001)]{todini2001}
Todini, P., \& Ferrara, A. 2001, MNRAS, 325, 726

\bibitem[Tominaga et al.(2008)]{tominaga2008}
Tominaga, N., et al. 2008, ApJ, 687, 1208

\bibitem[Whitelock et al.(1989)]{whitelock1989}
Whitelock, P. A., et al. 1989, MNRAS, 240, 7

\bibitem[Wooden et al.(1993)]{wooden1993}
Wooden, D. H., Rank, D. M., Bregman, J. D., Witteborn, F. C., 
Tielens, A. G. G. M., Cohen, M., Pinto, P. A., \& Axelrod, T. S. 
1993, ApJS, 88, 477

\bibitem[Xu et al.(1992)]{xu1992}
Xu, Y., McCray, R., Oliva, E., Randich, S. 1992, ApJ, 386, 181

\bibitem[Yaron \& Gal-Yam(2012)]{yaron2012}
Yaron, O \& Gal-Yan, A. 2012, PASP, 124, 668

\bibitem[Zhang et al.(2012)]{zhang2012}
Zhang, T., et al. 2012, AJ, 144, 131

\bibitem[Zubko et al.(1996)]{zubko1996}
Zubko, V. G., Mennella, V., Colangeli, L., \& Bussoletti, E. 
1996, MNRAS, 282, 1321

\end{thebibliography}
\end{document}